\documentclass[12pt]{article}

\usepackage{epsfig}
\usepackage{psfrag}
\usepackage{latexsym}
\usepackage[DIV13]{typearea}
\usepackage{amsmath}
\usepackage{amssymb}
\usepackage{amsfonts}
\usepackage{bbold}
\usepackage[footnotesize]{caption2}
\usepackage{graphicx}
\usepackage[center,footnotesize,hang]{subfigure}
\usepackage{url}
\usepackage{color}
\usepackage{accents}
\usepackage{cite}

\def\cL{{\cal L}}

\newcommand{\phit}{\varphi_T}
\newcommand{\phis}{\varphi_S}
\newcommand{\onep}{1^\prime}
\newcommand{\onepp}{1^{\prime\prime}}
\newcommand{\al}{\alpha}

\newcommand{\la}{\lambda}
\newcommand{\La}{\Lambda}
\newcommand{\om}{\omega}

\def\beq{\begin{equation}}
\def\eeq{\end{equation}}
\def\bea{\begin{eqnarray}}
\def\eea{\end{eqnarray}}
\def\ba{\begin{array}}
\def\ea{\end{array}}
\def\baq{\beq\ba{rcl}}
\def\eaq{\ea\eeq}
\def\bet{\begin{tabular}}
\def\eet{\end{tabular}}
\def\bes{\begin{subequations}\bea}
\def\ees{\eea\end{subequations}}
\DeclareMathOperator{\Tr}{Tr}
\newcommand{\mean}[1]{\langle#1\rangle}

\newcommand{\unity}{\mathbb{1}}
\newcommand{\eq}[1]{eq. (\ref{#1})}
\def\D{\mathrm{d}}
\def\dd{\displaystyle}
\def\nn{\nonumber}
\def\diag{\mathrm{diag}}

\catcode`@=11
\def\marginnote#1{}
\newcount\hour
\newcount\minute
\newtoks\amorpm
\hour=\time\divide\hour by60
\minute=\time{\multiply\hour by60 \global\advance\minute by-\hour}
\edef\standardtime{{\ifnum\hour<12 \global\amorpm={am}%
        \else\global\amorpm={pm}\advance\hour by-12 \fi
        \ifnum\hour=0 \hour=12 \fi
        \number\hour:\ifnum\minute<10 0\fi\number\minute\the\amorpm}}
\edef\militarytime{\number\hour:\ifnum\minute<10 0\fi\number\minute}
\def\draftlabel#1{{\@bsphack\if@filesw {\let\thepage\relax
   \xdef\@gtempa{\write\@auxout{\string
      \newlabel{#1}{{\@currentlabel}{\thepage}}}}}\@gtempa
   \if@nobreak \ifvmode\nobreak\fi\fi\fi\@esphack}
        \gdef\@eqnlabel{#1}}
\def\@eqnlabel{}
\def\@vacuum{}
\def\draftmarginnote#1{\marginpar{\raggedright\scriptsize\tt#1}}
\def\draft{\oddsidemargin 0.0truein
        \def\@oddfoot{\sl preliminary draft \hfil
        \rm\thepage\hfil\sl\today\quad\militarytime}
        \let\@evenfoot\@oddfoot \overfullrule 3pt
        \let\label=\draftlabel
        \let\marginnote=\draftmarginnote
   \def\@eqnnum{(\theequation)\rlap{\kern\marginparsep\tt\@eqnlabel}%
\global\let\@eqnlabel\@vacuum}  }
\catcode`@=12
\begin{document}
\begin{titlepage}
\vspace*{-1cm}
\phantom{hep-ph/***}
\hfill{DFPD-09/TH/20}\\
\vskip 2.5cm
\begin{center}
{\Large\bf Running Effects on Lepton Mixing Angles\\
\vskip .3cm
in Flavour Models with Type I Seesaw}
\end{center}
\vskip 0.5  cm
\begin{center}
{\large Yin Lin}\footnote{e-mail address: yin.lin@pd.infn.it},
{\large Luca Merlo}\footnote{e-mail address: merlo@pd.infn.it} and
{\large Alessio Paris}\footnote{e-mail address: paris@pd.infn.it}
\\
\vskip .2cm
Dipartimento di Fisica `G.~Galilei', Universit\`a di Padova
\\
INFN, Sezione di Padova, Via Marzolo~8, I-35131 Padova, Italy
\\
\end{center}
\vskip 0.2cm
\begin{abstract}
\noindent
We study renormalization group running effects on neutrino mixing patterns
when a (type I) seesaw model is implemented by suitable flavour symmetries.
We are particularly interested in mass-independent mixing patterns
to which the widely studied tribimaximal mixing pattern belongs.
In this class of flavour models, the running contribution from neutrino Yukawa coupling,
which is generally dominant at energies above the seesaw threshold,
can be absorbed by a small shift on neutrino mass
eigenvalues leaving mixing angles unchanged.
Consequently, in the whole running energy range,
the change in mixing angles is due to the contribution
coming from charged lepton sector.
Subsequently, we analyze in detail these effects in an explicit flavour model for tribimaximal
neutrino mixing based on an $A_4$ discrete symmetry group.
We find that for normally ordered light neutrinos,
the tribimaximal prediction is essentially stable under renormalization group evolution.
On the other hand, in the case of inverted hierarchy, the deviation of the solar angle
from its TB value can be large depending on mass degeneracy.
\end{abstract}
\end{titlepage}
\setcounter{footnote}{0}
\vskip2truecm

%%%%%%%%%%%%%%%%%%%%%%%%%%%%%%%%%%%%%%%%%%%%%%%%%%%%%%%%%%%%%%%%%%%%%%%%%%%%%%%%%%%%%%%%%%%%%%%%%%%%%%%%%%%%%%%%%%%%%%%%
%%%%%%%%%%%%%%%%%%%%%%%%%%%%%%%%%%%%%%%%%%%%%%%       1. INTRODUCTION      %%%%%%%%%%%%%%%%%%%%%%%%%%%%%%%%%%%%%%%%%%%%%
%%%%%%%%%%%%%%%%%%%%%%%%%%%%%%%%%%%%%%%%%%%%%%%%%%%%%%%%%%%%%%%%%%%%%%%%%%%%%%%%%%%%%%%%%%%%%%%%%%%%%%%%%%%%%%%%%%%%%%%%

\section{Introduction}

The seesaw mechanism \cite{SeeSaw} explains the lightness of left-handed (LH) neutrinos in a simple and elegant way. However, the seesaw models usually contain many more parameters
than low energy observables. In order to economically describe the neutrino oscillations,
new ingredients are needed: the introduction of a horizontal symmetry could improve the situation and indeed many examples of this kind are present in literature.
A common feature of these models is to provide a description of neutrino masses and mixings only at a very high energy scale. On the other hand, for a comparison of experimental results
with the high energy predictions from flavour symmetries,
it is important to evolve the observables to low energies through
renormalization group (RG) running.
In general the deviations from high energy values due to the RG running
consist in minor corrections, but the future improvements of neutrino experiments could hopefully
bring the precision down to these small quantities.

In the MSSM context, the RG effect on the light neutrino mass operator $m_\nu$
in the leading Log approximation can approximately be parametrized as
\beq
m_{\nu\,\text{(lower Energy)}}=I_U J^T_e J_\nu^Tm_{\nu\,\text{(higher Energy)}} J_\nu J_e
\nn
\eeq
where $I_U$ is a universal contribution, $J_e$, that is proportional to $Y^\dagger_e Y_e$
with $Y_e$ the charged Yukawa coupling, is always flavour dependent
and, $J_\nu$ may depend on flavour or not.
The running contributions to $m_\nu$ from $J_e$ are similar to those below the lightest
right-handed (RH) neutrino mass and it is as known well under control.
In fact large RG effects can be expected only in the case of a degenerate light neutrino
spectrum or for large values of $\tan \beta$ \cite{RunningNoSeeSaw}.
The running contribution from $J_\nu$,
however, is generally more complicated depending non-trivially on the neutrino Yukawa coupling
$Y_\nu$. Moreover, the last contribution can even be the dominant one
since $Y_\nu$ is expected to be of order one \cite{RunningSeeSaw}.
Similar conclusions can be driven in the SM context, when assuming a flavour origin of the neutrino mass term $m_\nu$.
In seesaw models implemented by flavour symmetries,
$Y_\nu$ is usually subject to constraints in order to efficiently describe the observed neutrino
mixing structure. We should expect that these constraints
have also some important impacts on running effects.
In this paper, we will show that, under quite general assumptions,
flavour symmetries imply a $J_\nu$ contribution which has not effects on
mixing angles. As a consequence even including energy ranges above and between the
RH neutrino scales, the contribution from the charged lepton Yukawa coupling
$J_e$ on mixing patterns dominates.

In the first part of the paper,
we will describe, in a very general context, two kinds of interesting constraints
on $Y_\nu$ from flavour symmetries and then analyze their impact on running effects.
We first consider flavour models in which $Y_\nu$ is proportional to a unitary matrix.
It is the case, for example, when the RH singlet neutrinos or the charged leptons
are in a irreducible representation of the flavour group $G_f$.

Then we extend this constraint to a more general class of flavour models in which the
mixing textures are independent from the mass eigenstates.
As a general result, we find that in this class of models,
the effect of $J_\nu$ can always be absorbed by a small shift on neutrino mass
eigenvalues and the mixing angles remain unchanged.
This conclusion is, in particular, independent both from the specific mixing pattern implied
by the flavour symmetry and the basis where we are working.

Mass-independent mixing textures usually exhibit
a underlying discrete symmetry nature.
For example, the tribimaximal (TB) mixing pattern  \cite{HPS} which
is defined by
\beq
U_{TB}=\left(
\begin{array}{ccc}
\sqrt{2/3}& 1/\sqrt{3}& 0\\
-1/\sqrt{6}& 1/\sqrt{3}& -1/\sqrt{2}\\
-1/\sqrt{6}& 1/\sqrt{3}& +1/\sqrt{2}
\end{array}
\right)~,
\label{UTB}
\eeq
belongs to this class of mixing textures.
TB mixing pattern, widely studied in the last years, provides a
very simple first order description of the existing oscillation
data \cite{Data, FogliIndication, MaltoniIndication}.
In fact, it has been realized that the TB mixing matrix of eq. (\ref{UTB})
can naturally arise as the result of a particular vacuum alignment of scalars
that break spontaneously certain discrete flavour symmetries.
A class of economical flavour models which can naturally
explains TB mixing is based on $A_4$ discrete symmetry \cite{AF_Papers} \cite{Lin_Papers} \cite{A4I,A4II}.
Subsequently the symmetry group $A_4$ has also been
extended to the group $T'$ \cite{Tp_Papers} and $S_4$ \cite{S4_Papers,BMM_SS} to cover a reasonable
description also for quarks.
Another important historical example of mass-independent mixing
scheme is the bimaximal (BM) mixing
(see \cite{BM} for a recent revival of BM in the context
of the discrete symmetry $S_4$ and for an up-to-date list of references).
The running effects on BM \cite{Running+BM} and TB \cite{Running+TB}
mixing patterns have already been studied in literature without,
however, consider an explicit realization based on flavour symmetry.

Then, as an explicit example, in the second part of the paper, we describe in detail the RG effects
on the TB mixing texture in the lepton flavour model proposed by Altarelli-Feruglio (AF)
\cite{AF_Papers}.  At leading order, this model contains less parameters than the case of a
general TB pattern considered in \cite{Running+TB}. Then our result may not
a priori reproduce the same results obtain in \cite{Running+TB}.
In our analysis, the RG corrections to mixing angles and the phases
are discussed as functions of the lightest neutrino mass and the type of spectrum.
The analysis has been performed both in the Standard Model (SM) framework
and the its minimal SUSY extension (MSSM).

The outline of the paper is organized as follows.
In section 2, we analytically discuss the RG effects on the
neutrino mass operator $m_\nu$ in a type I seesaw model.
In section 3, we characterize two general classes of flavour models in which
$J_\nu$ has no effects on mixing angles.
Then we turn to RG effects from $J_e$ on TB mixing pattern.
In section 4 we introduce the AF model and show its main features
including also the next-to-leading (NLO) contributions coming from
higher dimensional operators.
In section 5, a more detailed numerical analysis on the impact of RG running in
the AF model is given. The result has been compared also with the NLO contributions.
In the end in section 6 we conclude summarizing our main results.

%%%%%%%%%%%%%%%%%%%%%%%%%%%%%%%%%%%%%%%%%%%%%%%%%%%%%%%%%%%%%%%%%%%%%%%%%%%%%%%%%%%%%%%%%%%%%%%%%%%%%%%%%%%%%%%%%%%%%%%%
%%%%%%%%%%%%%%%%%%%%%%%%%%       2. RG effects on neutrino mass operator ${\bf m_\nu}$   %%%%%%%%%%%%%
%%%%%%%%%%%%%%%%%%%%%%%%%%%%%%%%%%%%%%%%%%%%%%%%%%%%%%%%%%%%%%%%%%%%%%%%%%%%%%%%%%%%%%%%%%%%%%%%%%%%%%%%%%%%%%%%%%%%%%%%

\section{RG effects on neutrino mass operator ${\bf m_\nu}$}
\label{Sec_Running}

In this section we begin to analyze, in a general context, the RG equations for neutrino masses below and above the seesaw threshold, both in the SM and in MSSM extended with three right-handed neutrinos.
For definiteness, we consider the following lagrangian
\beq
{\cal L}=e^{c\,T} Y_e H^\dag \ell+ \nu^{c\,T} Y_\nu \widetilde H^\dag \ell + \nu^{c\,T} M_R \nu^c +h.c.
\eeq
where $\ell$ are the LH lepton doublets, $e^c$ the RH charged lepton singlets and $H (\widetilde H\equiv i\sigma_2 H^*)$ is the Higgs doublet. In the supersymmetric case, this lagrangian should be identified to a superpotential where $H (\tilde H)$ is replaced by $h_d (h_u)$ and all the fields are instead supermultiplets. In what follows we concentrate only on the SM particles, postponing the study of their supersymmetric partners elsewhere: for this reason in our notation a chiral superfield and its $R$-parity even component are denoted by the same letter.

Given the heavy Majorana and the Dirac neutrino mass matrices, $M_R$ and \mbox{$m_D=Y_\nu v/\sqrt2$} respectively, the light neutrino one is obtained from block-diagonalising the complete $6\times6$ neutrino mass matrix:
\beq
m_\nu=-\dfrac{v^2}{2}Y_\nu^TM_R^{-1}Y_\nu\;,
\label{EqSeeSaw}
\eeq
where $v$ refers to the VEV of the Higgs field, $\mean{H}\equiv v/\sqrt2$ ($v\approx246$ GeV). The equivalent relation in the supersymmetric case is achieved by replacing $v$ with $v_u$. In our notation the VEV of the Higgs fields $h_u$ and $h_d$ are given by $\mean{h_{u,d}}\equiv v_{u,d}/\sqrt2$, with $\sqrt{v_u^2+v_d^2}=v$. The matrix $m_\nu$ is modified by quantum corrections according to the renormalization group equations (RGEs) widely studied in literature \cite{RunningSeeSaw}. For completeness, in
Appendix A, we report the full RGEs for all the interested quantities in the running. Here, we explicitly give RGEs  only for the Yukawa couplings $Y_e$, $Y_\nu$ and the RH neutrino mass $M_R$,
relevant for the analysis on this section: in the SM context they are written as
\beq
\hspace{-3mm}
\ba{lcl}
16\pi^2 \dfrac{d}{d t}Y_e &\!\!=\!\! & Y_e\left\{\dfrac{3}{2}Y_e^\dagger Y_e -\dfrac{3}{2} Y_\nu ^\dagger  Y_\nu + \Tr\left[ 3Y_u^\dagger Y_u + 3Y_d^\dagger Y_d + Y_\nu^\dagger Y_\nu + Y_e^\dagger Y_e \right] - \dfrac{9}{4}g_1^2 - \dfrac{9}{4} g_2^2 \right\}\;,\\[3mm]
16\pi^2 \dfrac{d}{d t}Y_\nu &\!\!=\!\!& Y_\nu \left\{\dfrac{3}{2} Y^\dagger_\nu Y_\nu -\dfrac{3}{2} Y_e^\dagger Y_e + \Tr\left[ 3Y_u^\dagger Y_u + 3Y_d^\dagger Y_d + Y_\nu^\dagger Y_\nu + Y_e^\dagger Y_e \right] - \dfrac{9}{20} g_1^2 - \dfrac{9}{4} g_2^2 \right\}\;,\\[3mm]
16\pi^2 \dfrac{d}{d t} M_R &\!\!=\!\!& \left(\,Y_\nu Y^\dagger_\nu \right)\, M_R + \,M_R\,\left(Y_\nu Y^\dagger_\nu \right)^T\;,
\ea
\label{EqRGESM}
\eeq
while in the MSSM they are slightly different,
\beq
\ba{lcl}
16\pi^2 \dfrac{d}{d t}Y_e & = & Y_e\left\{3Y_e^\dagger Y_e + Y_\nu ^\dagger  Y_\nu + \Tr\left[ 3Y_d^\dagger Y_d + Y_e^\dagger Y_e \right] - \dfrac{9}{5}g_1^2 - 3g_2^2 \right\}\;,\\[3mm]
16\pi^2 \dfrac{d}{d t}Y_\nu &=& Y_\nu \left\{ 3 Y^\dagger_\nu Y_\nu + Y_e^\dagger Y_e + \Tr\Big[ 3Y_u^\dagger Y_u + Y^{\dagger}_\nu Y_\nu \Big] - \dfrac{3}{5} g_1^2 - 3 g_2^2 \right\}\;,\\[3mm]
16\pi^2 \dfrac{d}{d t} M_R &=& 2\left(\,Y_\nu Y^\dagger_\nu \right)\, M_R + 2\,M_R\,\left(Y_\nu Y^\dagger_\nu \right)^T\;,
\ea
\label{EqRGEMSSM}
\eeq
where  $t:=\ln(\mu/\mu_0)$ and $Y_{u(d)}$ is the Yukawa coupling for the up- (down-) quarks.\footnote{In the GUT normalization, such that $g_2=g$ and $g_1=\sqrt{5/3} g'$.}
From these RGEs, one can obtain the RG evolution for the
running composite operator $m_\nu (\mu)$ defined in eq.~(\ref{EqSeeSaw}).
In order to analytically study the change of $m_\nu(\mu)$ from high to low energy,
it is useful to work in the basis in which the Majorana neutrino mass is diagonal and real, $\hat{M}_R = {\rm diag}(M_S, M_M, M_L)$. The mass eigenvalues can be ordered as $M_S < M_M < M_L$. Furthermore, we can divide the RG effects in three distinct energy ranges: from the cutoff $\Lambda$ of the theory down to $M_L$, the mass of the heaviest RH neutrino; from $M_L$ down to $M_S$, the mass of the lightest RH neutrino; below $M_S$ down to $\lambda$, which can be either $m_Z$, considered as the electroweak scale, or $m_\mathrm{SUSY}$, the average energy scale for the supersymmetric particles.

\begin{description}
\item[$\mathbf{\La_f\longrightarrow M_L}$.]
Above the highest seesaw scale the three RH neutrinos are all active and the dependence of the effective light neutrino mass matrix from the renormalization scale $\mu$ is given by mean of the $\mu-$dependence of $Y_\nu$ and $M_R$:
\begin{equation}
m_\nu(\mu)\, =\, -\dfrac{v^2}{2} \, Y_\nu^T(\mu)\, M_R^{-1}(\mu) \,Y_\nu(\mu) \;.
\label{effnumass1}
\end{equation}
Then from the RGEs in eqs. (\ref{EqRGESM}, \ref{EqRGEMSSM}), it is not difficult to see that the evolution of the composite operator $m_\nu$ is given by:
\begin{equation}
16\pi^2 \, \frac{\D m_\nu}{\D t} =\Big(C_eY_e^\dagger Y_e + C_\nu Y_\nu^\dagger Y_\nu\Big)^T \, m_\nu +m_\nu \, \Big(C_e Y_e^\dagger Y_e + C_\nu Y_\nu^\dagger Y_\nu\Big) + \bar\alpha \, m_\nu
\label{betanu1}
\end{equation}
with
\beq
\ba{ll}
C_e=\,-\dfrac{3}{2}\;,\qquad C_\nu=\dfrac{1}{2}&\qquad\text{in the SM}\\[5mm]
C_e=\,C_\nu=1&\qquad\text{in the MSSM}
\ea
\eeq
and
\beq
\ba{ccl}
\bar\alpha_{SM}&=&2\Tr\left[ 3Y_u^\dagger Y_u + 3Y_d^\dagger Y_d + Y_\nu^\dagger Y_\nu + Y_e^\dagger Y_e \right] - \dfrac{9}{10} g_1^2 - \dfrac{9}{2} g_2^2\\[5mm]
\bar\alpha_{MSSM}&=&2\Tr\Big[ 3Y_u^\dagger Y_u + Y^{\dagger}_\nu Y_\nu \Big] - \dfrac{6}{5} g_1^2 - 6 g_2^2\;.
\ea
\eeq

\item[$\mathbf{M_L\longrightarrow M_S}$.]
The effective neutrino mass matrix $m_\nu$ below the highest seesaw scale
can be obtained by sequentially integrating out $\nu^c_n$ with $n=L,M,S$:
\begin{equation}\label{effnumass2}
m_\nu \,=\,-\dfrac{v^2}{4} \left(\,\accentset{(n)}{\kappa}+ 2 \accentset{(n)}{Y}_\nu^T\accentset{(n)}{M_R}^{-1}\accentset{(n)}{Y}_\nu\right)
\end{equation}
where $\accentset{(n)}{\kappa}$ is the coefficient of the effective neutrino mass operator $ \widetilde H^\dag \ell\widetilde H^\dag\ell$. From the (tree-level) matching condition, it is given by
\begin{equation}
\accentset{(n)}{\kappa}_{ij}\,=\,2(Y_\nu^T)_{in}\,M^{-1}_n\,(Y_\nu)_{nj}\;,
\label{kn}
\end{equation}
which is imposed at $\mu=M_n$.
At $M_L$, the $2\times3$ Yukawa matrix $\accentset{(L)}{Y}_\nu$ is obtained by simply removing the
$L$-th row of $Y_\nu$ and the $2\times2$ mass matrix $\accentset{(L)}{M_R}$
is found from $M_R$ by removing the $L$-th row and $L$-th column. Further decreasing the energy scale down to $M_M$, $\accentset{(M)}{Y}_\nu$ is a single-row matrix, obtained by removing the $M$-th row from $\accentset{(L)}{Y}_\nu$, and
$\accentset{(M)}{M_R}$ consists of a single parameter, found by removing the $M$-th row and $M$-th column from $\accentset{(L)}{M}_R$. Finally at $M_S$, $\accentset{(S)}{Y}_\nu$ and $\accentset{(S)}{M}_R$ are vanishing.

In the SM, the two parts which define $m_\nu$ in eq. (\ref{effnumass2}) evolve in different ways. We can summarize the corresponding RGEs as follows:
\beq
16\pi^2 \, \frac{\D \accentset{(n)}{X}}{\D t} = \left( \dfrac{1}{2}\accentset{(n)}{Y}_\nu^\dagger \accentset{(n)}{Y}_\nu - \dfrac{3}{2}Y_e^\dagger Y_e \right)^T \accentset{(n)}{X} + \accentset{(n)}{X} \left( \dfrac{1}{2}\accentset{(n)}{Y}_\nu^\dagger \accentset{(n)}{Y}_\nu - \dfrac{3}{2}Y_e^\dagger Y_e \right) +  \accentset{(n)}{\bar\alpha}_X \accentset{(n)}{X}
\eeq
where
\beq
\ba{ccl}
\accentset{(n)}{\bar\alpha}_\kappa &=& 2\Tr\left[ 3Y_u^\dagger Y_u + 3Y_d^\dagger Y_d + \accentset{(n)}{Y}_\nu^\dagger \accentset{(n)}{Y}_\nu + Y_e^\dagger Y_e \right] -3 g_2^2 + \lambda_H\\[5mm]
\accentset{(n)}{\bar\alpha}_{Y_\nu^TM_R^{-1}Y_\nu} &=& 2\Tr\left[ 3Y_u^\dagger Y_u + 3Y_d^\dagger Y_d + \accentset{(n)}{Y}_\nu^\dagger \accentset{(n)}{Y}_\nu + Y_e^\dagger Y_e \right]-\dfrac{9}{10} g_1^2 - \dfrac{9}{2} g_2^2\;,
\ea
\eeq
with $\lambda_H$ the Higgs self-coupling.\footnote{We use the convention that the Higgs self-interaction term in the
Lagrangian is $-\lambda_H (H^\dagger H)^2/4$.}

In MSSM the running of $\accentset{(n)}{\kappa}$ and of $\accentset{(n)}{Y}_\nu^T\accentset{(n)}{M_R}^{-1}\accentset{(n)}{Y}_\nu$ is the same and therefore we can write
\beq
16 \pi^2 \,
\frac{\D m_\nu} {\D t}\,=\,\left(Y_e^\dagger Y_e +\accentset{(n)}{Y}_\nu^\dagger\accentset{(n)}{Y}_\nu\right)^T m_\nu + m_\nu \left( Y_e^\dagger Y_e +  \accentset{(n)}{Y}_\nu^\dagger \accentset{(n)}{Y}_\nu\right) +  \accentset{(n)}{\bar\alpha}m_\nu \; ,
\label{betanu2}
\eeq
where
\beq
\accentset{(n)}{\bar\alpha} = 2\Tr\left[ 3Y_u^\dagger Y_u + \accentset{(n)}{Y}_\nu^\dagger \accentset{(n)}{Y}_\nu \right] -\dfrac{6}{5} g_1^2 - 6 g_2^2\;.
\eeq

\item[$\mathbf{M_S\longrightarrow\la}$.] For energy range below the mass scale of the lightest RH neutrino,
all the $\nu^c_n$ are integrated out and $\accentset{(S)}{Y}_\nu$ and $\accentset{(S)}{M}_R$ vanish. In the right-hand side of eq. (\ref{effnumass2}) only the term $\accentset{(S)}{\kappa}$ is not vanishing and in this case the composite operator $m_\nu$ evolves as:
\begin{equation}
16\pi^2 \, \dfrac{\D m_\nu}{\D t} = \Big(C_eY_e^\dagger Y_e\Big)^T \, m_\nu + m_\nu \, \Big(C_eY_e^\dagger Y_e\Big) + \accentset{(S)}{\bar\alpha} \, m_\nu
\label{betanu3}
\end{equation}
with
\begin{equation}
\ba{ccl}
\accentset{(S)}{\bar\alpha}_{SM}&=& 2\Tr\left[ 3Y_u^\dagger Y_u + 3Y_d^\dagger Y_d + Y_e^\dagger Y_e \right] -3 g_2^2 + \lambda_H\\[5mm]
\accentset{(S)}{\bar\alpha}_{MSSM}&=& 6\Tr\Big[Y_u^\dagger Y_u \Big] - \dfrac{6}{5} g_1^2 - 6 g_2^2\;.
\ea
\end{equation}
\end{description}

%
%%%%%%%%%%%%%%%%%%%%%%%%%%%%%%%%%%%%%%%%%%%%%%%       2.1 Analytical approximation to RG evolution of $m_\nu$        %%%%%%%%%%%%%%%%%%%%%%
%

\subsection{Analytical approximation to RG evolution of ${\bf m_\nu}$}
\label{Subsec_Analytical_RGE}

Now we analytically solve the RG equations for $m_\nu$ in the leading Log approximation.
All the Yukawa couplings $Y_i^\dagger Y_i$ for $i=\nu,e,u,d$ are valuated at their initial value
at the cutoff $\La_f$. Furthermore we will keep only the leading contributions from each
$Y_i^\dagger Y_i$ term, for $i=e,u,d$, i.e. $|y_\tau|^2$, $|y_t|^2$ and $|y_b|^2$ respectively.
The RG corrections to these quantities would contribute to the final result
as sub-leading effects and we can safely neglect them in the analytical estimate.

In the MSSM context, the general solution to eqs.~(\ref{betanu1}), (\ref{betanu2}) and (\ref{betanu3})
have all the same structure, which is approximately given by
\beq
m_{\nu\,\text{(lower Energy)}} \approx I_U J_e^T J_\nu^Tm_{\nu\,\text{(higher Energy)}} J_\nu J_e
\label{generalsol}
\eeq
where $I_U$, $J_e$ and $J_\nu$ are all exponentials of integrals containing loop suppressing factors and as a result they are close to $\unity$. Note that $I_U$ is a universal contribution defined as
\beq
I_U=\exp\left[-\dfrac{1} {16\pi^2}\int\accentset{(n)}{\bar\alpha}~\D t \right]
\label{I}
\eeq
where the integral runs between two subsequent energy scales and we have extended the definition of $\accentset{(n)}{\bar\alpha}$
by identifying $\accentset{(\Lambda)}{\bar\alpha} \equiv  \bar\alpha$ in order to include
the range from $\Lambda$ down to $M_L$.
$J_e$ is the contribution from charged lepton Yukawa couplings which
is always flavour-dependent and is given by\footnote{In eq. (\ref{EqJeGeneral}), the combination $Y_e^\dagger Y_e$ should enter with $\accentset{(n)}{Y}_e$ instead of $Y_e$, as one can see from the RGEs in Appendix A. In our approximation, however, they coincide.}
\beq
J_{e}=\exp\left[-\dfrac{1}{16\pi^2}\int Y_e^\dagger Y_e
~\D t \right]\;.
\label{EqJeGeneral}
\eeq
Finally, $J_\nu$ is the contribution from the neutrino Yukawa coupling
\beq
J_{\nu}=\exp\left[-\dfrac{1}{16\pi^2}\int\accentset{(n)}{Y}_{\nu}^\dagger \accentset{(n)}{Y}_{\nu} ~\D t \right]\;,
\eeq
where also here we have extended the definition of $\accentset{(n)}{Y}_{\nu}$ by identifying $\accentset{(\La)}{Y}_{\nu}$ with $Y_\nu$ in order to include the range between $\Lambda$ and $M_L$.
Differently from $J_e$, $J_\nu$ can be flavour-dependent or not.

In the SM context, the RG effect does not factorize, due to the different RG evolution of $\accentset{(n)}{\kappa}$ and $\accentset{(n)}{Y}_\nu^T \accentset{(n)}{M}_R^{-1} \accentset{(n)}{Y}_\nu$ between the seesaw mass thresholds. However eq. (\ref{generalsol}) applies also to the SM context when $m_\nu$ is a result of a flavour symmetry: in this case, by a suitable redefinition of the mass eigenvalues, the sum $\accentset{(n)}{\kappa}+\accentset{(n)}{Y}_\nu^T \accentset{(n)}{M}_R^{-1} \accentset{(n)}{Y}_\nu$ after the RG evolution has exactly the same flavour structure of $m_{\nu\,\text{(higher Energy)}}$. For the purposes of the present discussion we simply assume that eq. (\ref{generalsol}) is valid also in the SM context and an explicit example will be proposed in section \ref{RGcoefficients}.

Expanded $J_e$ and $J_\nu$ in Taylor series and summing
up (\ref{generalsol}) on several energy ranges
one can approximately calculate the neutrino mass at low energy as
\beq
m_{\nu(\lambda)}\simeq I_U\left(m_{\nu(\Lambda)}+
\Delta m^{(J_e)}_\nu + \Delta m^{(J_\nu)}_\nu\right)\;,
\label{generalsol2}
\eeq
where the low energy scale $\lambda$ is $m_Z$ in the case of SM and $m_{\rm SUSY}$
for MSSM. The explicit form of the universal part $I_U$ is given by:
\bea
&&\begin{split}
\hspace{-7mm}
I_U^{\rm SM}\;=&\;\unity\;\times\;\exp\Bigg[-\dfrac{1}{16\pi^2}\Bigg[\left(-\dfrac{9}{10}g_1^2-\dfrac{9}{2}g_2^2+ 6|y_t|^2\right)\ln\dfrac{\La_f}{m_Z}+ \left(\dfrac{9}{10}g_1^2+\dfrac{3}{2}g_2^2+\lambda_H\right)\ln\dfrac{M_S}{m_Z}+\\[3mm]
&\hspace{3cm}+y^2\left(2\ln\dfrac{M_M}{M_S}+ 4\ln\dfrac{M_L}{M_M}+7\ln\dfrac{\La_f}{M_L}\right)\Bigg]\Bigg]\;,
\end{split}\nn\\
\label{IUfinale}\\[-3mm]
&&\begin{split}
\hspace{-7mm}
I_U^{\rm MSSM}\;=&\;\unity\;\times\;\exp\Bigg[-\dfrac{1}{16\pi^2}\Bigg[\left(-\dfrac{6}{5}g_1^2-6g_2^2+ 6|y_t|^2\right)\ln\dfrac{\La_f}{m_\mathrm{SUSY}}+\\[3mm]
&\hspace{3cm}+y^2\left(2\ln\dfrac{M_M}{M_S}+ 4\ln\dfrac{M_L}{M_M}+8\ln\dfrac{\La_f}{M_L}\right)\Bigg]\Bigg]\;.\nn
\end{split}
\eea
$\Delta m^{(J_e)}_\nu$ is the the contribution from $J_e$ and can easily be calculated as:
\beq
\Delta m^{(J_e)}_\nu = m_{\nu (\Lambda)} ~ {\rm diag} (0, 0, \Delta_\tau)+{\rm diag} (0, 0, \Delta_\tau) m_{\nu (\Lambda)}
\label{Je}
\eeq
where the small parameter $\Delta_\tau$ is given by
\beq
\ba{ccll}
\Delta_\tau&\equiv&-\dfrac{3m^2_\tau}{16\pi^2 v^2}\ln\dfrac{\Lambda}{m_Z}&\quad\text{in the SM}\\[3mm]
\Delta_\tau&\equiv&\dfrac{m^2_\tau}{8\pi^2 v^2} (1+ \tan^2 \beta) \ln\dfrac{\Lambda}{m_\mathrm{SUSY}}&\quad\text{in the MSSM}\;
\ea
\label{DeltaTau}
\eeq
where $\tan\beta$ is the ratio between the VEVs of the neutral spin zero components of $h_u$ and $h_d$, the two doublets responsible for electroweak symmetry breaking in the MSSM.
On the other hand, the contribution from $J_\nu$, $\Delta m^{(J_\nu)}_\nu$, non trivially depends on the neutrino Yukawa coupling
$Y_\nu$ which cannot be determined by low energy observables without additional ingredients.
In section \ref{Sec_FlavourSym_RGE}, we will analyze strong impacts of
the flavour symmetries on $J_\nu$. But before proceeding,
we can first give a naive estimate of various running contributions to neutrino mass.

%
%%%%%%%%%%%%%%%%%%%%%%%%%%%%%%%%%%%%%%%%%%%%%%%       2.2 Naive numerical estimate        %%%%%%%%%%%%%%%%%%%%%%
%

\subsection{Naive numerical estimate}
\label{Sec_NaiveNumericalEstimate}

Here we numerically estimate the contributions of running, encoded in
$\Delta m_\nu$, above the seesaw scales
($\Delta m_{\nu(\mathrm{high})}$), those between the seesaw scales
($\Delta m_{\nu(\mathrm{seesaw})}$)
and below the seesaw scales ($\Delta m_{\nu(\mathrm{low})}$) on the light neutrino mass matrix.
One should expect that the mixing angles will be affected by the same quantities.
We assume that flavour symmetries have not effects on $Y_\nu$ and then,
without cancellations,
$$ Y_\nu^\dagger Y_\nu \sim
\accentset{(n)}{Y}_\nu^\dagger\accentset{(n)}{Y}_\nu
= \mathcal{O}(1)~.$$
As a result, both in the energy range above and between the seesaw scales,
the running contributions from $Y_\nu$ which are all encoded in $J_\nu$
would dominate over the one from $Y_e$, encoded in $J_e$~.
Summing up all the two contributions we have
\beq
|\Delta m_{\nu(\mathrm{high})}| + |\Delta m_{\nu(\mathrm{seesaw})}|
\sim |\Delta m_\nu^{(J_\nu)}|
\sim\dfrac{1}{16\pi^2}\ln\dfrac{\Lambda}{M_S}
\label{seesawnaive}
\eeq
both in the SM and in the MSSM contexts. Below the mass
of the lightest RH neutrino $M_S$
the contribution $J_\nu$ is absent and we should roughly obtain
\beq
\ba{rcll}
|\Delta m_{\nu(\mathrm{low})}|&\sim&\dfrac{3m^2_\tau}{16\pi^2 v^2}\ln\dfrac{M_S}{m_Z}&\quad\text{in the SM}\\[3mm]
|\Delta m_{\nu(\mathrm{low})}|&\sim&\dfrac{m^2_\tau}{8\pi^2 v^2}(1+\tan^2\beta)\ln\dfrac{M_S}{m_\mathrm{SUSY}}&\quad\text{in the MSSM}\;.
\ea
\eeq
Now we can compare the two contributions $\Delta m_\nu^{(J_\nu)}$
and $\Delta m_{\nu (\rm low)}$. In general one assumes a seesaw scale between
$10^{12}$ GeV and $10^{14} $ GeV and a cutoff scale approximate to the
grand unified scale $\La_f\sim10^{16}$ GeV. For the numerical estimate we use
$M_S \sim 10^{12} $ GeV, $m_\mathrm{SUSY} \sim 10^3$ GeV and $m_Z\sim100$ GeV.
Then we obtain\footnote{We take the pole mass for the lepton $\tau$\cite{pdg2008}: $m_\tau=(1776.84\pm0.17)$ MeV.}
\beq
\ba{rcll}
\dfrac{|\Delta m_{\nu(\mathrm{low})}|}{|\Delta m^{(J_\nu)}_{\nu}|} & \sim &
\dfrac{3m^2_\tau}{v^2} \dfrac{8}{5}\sim2.5\times10^{-4}&\quad\text{in the SM}\\[3mm]
\dfrac{|\Delta m_{\nu(\mathrm{low})}|}{|\Delta m^{(J_\nu)}_{\nu}|} & \sim &
\dfrac{2m^2_\tau}{v^2} \dfrac{7}{5}(1+\tan^2\beta)
\sim1.5\times10^{-4}\tan^2\beta&\quad\text{in the MSSM}\;.
\ea
\eeq
We can conclude that both in the SM and the MSSM, the contribution from $J_\nu$
always dominates over $\Delta m_{\nu(\mathrm{low})}$ even for large
$\tan\beta$ (we consider $\tan\beta=60$ as the maximal value).

The previous naive estimate for RG effects on the neutrino mass operator, however,
cannot always be transfered to the change in mixing angles $\Delta \theta_{ij}$ for
a particular model building. Indeed, as we will see in a moment, quite frequently flavour symmetries imply a $J_\nu$ which is flavour-independent or has no effects on mixing angles. If it is the case, the situation changes drastically: even if $\Delta m_{\nu(\mathrm{low})}$ can still be dominated by $\Delta m_{\nu(\mathrm{seesaw})}$, $\Delta \theta_{ij(\mathrm{low})}$ should dominates over $\Delta \theta_{ij(\mathrm{seesaw})}$.

%%%%%%%%%%%%%%%%%%%%%%%%%%%%%%%%%%%%%%%%%%%%%%%%%%%%%%%%%%%%%%%%%%%%%%%%%%%%%%%%%%%%%%%%%%%%%%%%%%%%%%%%%%%%%%%%%%%%%%%%
%%%%%%%%%%%%%%%%%%%%%%%%%%       3. Flavour symmetries and RGE effects  %%%%%%%%%%%%%
%%%%%%%%%%%%%%%%%%%%%%%%%%%%%%%%%%%%%%%%%%%%%%%%%%%%%%%%%%%%%%%%%%%%%%%%%%%%%%%%%%%%%%%%%%%%%%%%%%%%%%%%%%%%%%%%%%%%%%%%

\section{Flavour symmetries and RGE effects}
\label{Sec_FlavourSym_RGE}

In the present section, we will apply the general results of the RG evolution of the
neutrino mass operator $m_\nu$ to models beyond the Standard Model,
where a flavour symmetry is added to the gauge group of the SM.
The main aim is to track some interesting connections between the running effects
and how the flavour symmetry is realized in nature.

In a given basis, $Y_e^\dagger Y_e$ and $m_\nu$ can be diagonalised by unitary matrices,
$U_e$ and $U_\nu$, respectively.
The lepton mixing matrix is given by $U_\mathrm{PMNS} = U_e^\dagger U_\nu$.
The analysis of how $U_\mathrm{PMNS}$
changes with the RG running has already extensively performed \cite{RunningNoSeeSaw} in the context of SM and MSSM. On the other hand, only few studies \cite{RunningSeeSaw} are present in literature considering the presence of additional RH neutrinos, which originate the type I seesaw mechanism. Here we develop a general RG analysis for seesaw models in which
the lepton mixing matrix $U_{\rm PMNS}$ is dictated by a flavour symmetry $G_f$.
It is a common feature in flavour model building that $G_f$ must be spontaneously broken in order to naturally describe fermion masses and mixings, as we will see in section \ref{Sec_AFmodel}. Here, we simply assume that $G_f$ is spontaneously broken by a set of flavon fields $\Phi$ at a very high scale. The symmetry group can be discrete or continuous, global or local (or even
a combination of them). Suppose that, at leading order, the neutrino mixing matrix is given
by $U_0$ which differs from $U_{\rm PMNS}$ by subleading contributions $\sim \langle \Phi \rangle
/ \Lambda_f$ where $\Lambda_f$ is the cutoff scale of the flavour symmetry $G_f$.
We will begin with some general assumptions on $U_0$ without however
specifying its form. Then we will move to specialize in a concrete case in which $U_0$
is given by the TB mixing pattern.

%
%%%%%%%%%%%%%%%%%%%%%%%%%%%%%%%%%%%%%%%%%%%%%%%       3.1 Running effects on neutrino mixing patterns       %%%%%%%%%%%%%%%%%%%%%%%
%

\subsection{Running effects on neutrino mixing patterns}

As described in Sec.~(\ref{Sec_Running}) the relevant running effects on $m_\nu$
are encoded in the combinations $Y^\dagger_e Y_e$ and $Y^\dagger_\nu Y_\nu$. Furthermore, we observe that a relevant contribution to the running of $Y^\dagger_e Y_e$ is encoded by $Y^\dagger_\nu Y_\nu$.
We perform the analysis in the basis in which the charged leptons are diagonal,
then at high energy we have
\beq Y^\dagger_e Y_e = {\rm diag} (m^2_e, m^2_\mu, m^2_\tau)\dfrac{2}{v^2}~~~.
\eeq
From now on, we will use $v$ in the notation of the SM and in order to convert
similar expressions to the MSSM, it is sufficient to substitute $v$ with $v_{u,d}$, when dealing with neutrinos or charged leptons, respectively.
Naturally, this simple form should change when evolving down to low energies.
The running effect of  $Y^\dagger_e Y_e$ on $m_\nu$ is of second order and we can safely forget it.
However it can generate a non trivial $U_e$ and consequently introduces additional corrections
to $U_{\rm PMNS}$. We will return to this effect in section \ref{Subsec_Chargedsector}.

Since flavour symmetries impose constraints on $Y_\nu$~, they should have some impacts also
on running effects. In this paper we are interested in two classes of constraints.
The first class is characterized by $Y_\nu$ proportional to a unitary matrix
and in the second one
we assume that $m_\nu$ can be exactly diagonalized by $U_0$ according to
\beq
\hat{m}_\nu = U^T_0 m_\nu U_0
\label{Diag_nu}
\eeq
where $\hat{m}_\nu = {\rm diag} (m_1, m_2, m_3)$ with $m_i$ positive and $U_0$ is a mass-independent mixing pattern enforced by the flavour symmetry $G_f$. Independently from
the way $G_f$ is broken, we will show that, in this second case, the neutrino
Yukawa coupling in the basis of diagonal RH Majorana neutrinos, which we indicate as $\hat{Y}_\nu$,
has the following simple form
\beq
\hat{Y}_\nu = i D\,U^\dagger_0
\label{General_hatY}
\eeq
where $D={\rm diag} (\pm \sqrt{2m_1M_1},\pm \sqrt{2m_2M_2},\pm \sqrt{2m_3M_3})/v$.

%%%%%%%%%%%%%%%
%%%%%%%%%%%%%%%
\subsubsection{${\bf Y_\nu^\dagger Y_\nu = Y_\nu Y_\nu ^\dagger \sim \unity}$}
\label{Subsec_UnitaryYnu}

As already pointed out, $Y^\dagger_\nu Y_\nu$ plays an important
role in determining the running effect on $m_\nu$. In the case of a
$Y_\nu$ proportional to a unitary matrix the study of RG evolutions becomes quite simple.
$Y_\nu^\dagger Y_\nu \sim \unity$ or $Y_\nu Y_\nu ^\dagger \sim \unity$
is quite frequent in the presence of a flavour symmetry. It is, for example, a consequence
of the first Schur's lemma when $\ell$ or $\nu^c$
transforms in a irreducible representation of the group $G_f$ \cite{BBFN_Lepto}.

Between the energy scales $\Lambda_f$
and $M_L$ one has $J_\nu \propto \unity$ then its contribution to $m_\nu$ is global.
Now we demonstrate the contribution from $J_\nu$ is universal also between
$M_L$ and $M_S$ if $Y_\nu^\dagger Y_\nu \sim \unity$.
We first consider the running effect on $\accentset{(n)}{\kappa}_{ij}$
defined in eq.~(\ref{kn}) which is proportional to
\beq
( \accentset{(n)}{Y}_\nu^\dagger \accentset{(n)}{Y}_\nu )^T\, \accentset{(n)}{\kappa}
~+~{\rm symmetrization}~. \nn
\eeq
In components, for the first term we have
\begin{equation}
2\sum_{m \ne l} \sum_k (Y^T_\nu)_{im}
 ( {Y}_\nu^* )_{mk}
(Y_\nu^T)_{kl}\,M^{-1}_l\,(Y_\nu)_{lj} =0 \nn
\end{equation}
where we have used the unitary condition $\sum_k ({Y}_\nu^* )_{mk}
(Y_\nu^T)_{kn} = \delta_{mn}$. As we expected $J_\nu$ has no non-global effects on $\accentset{(n)}{\kappa}$.
Now we move to analyze the second term in (\ref{effnumass2}). The running effect on this term
is proportional to
\beq
( \accentset{(n)} Y_\nu^\dagger \accentset{(n)}{Y}_\nu )^T\,
\accentset{(n)}{Y}_\nu^T\accentset{(n)}{M_R}^{-1}\accentset{(n)}{Y}_\nu
~+~{\rm symmetrization}~, \nn
\eeq
Similarly as before, using the unitary condition, we obtain in components
\beq
\sum_{m,p,q \ne n} \sum_k  (Y^T_\nu)_{im}
 ( {Y}_\nu^* )_{mk}
(Y_\nu^T)_{kp}
({M_R}^{-1})_{pq} ({Y}_\nu)_{qj}
= \sum_{p,q \ne n} (Y^T_\nu)_{ip}
({M_R}^{-1})_{pq} ({Y}_\nu)_{qj} ~~~, \nn
\eeq
which coincides with the starting observable, $\accentset{(n)}{Y}_\nu^T\accentset{(n)}{M_R}^{-1}\accentset{(n)}{Y}_\nu$. In the previous expression, the meaning of the apex $(n)$ is understood. A similar result is valid for the symmetrization part.
Then we conclude that $m_\nu$ does not change the flavour structure under $J_\nu$ if
$Y_\nu$ is proportional to a unitary matrix. This result in completely general
and holds in any basis, not only in the hatted one.
In this case, the only flavour-dependent RG contribution to $m_\nu$
is encoded in $J_e$.

%%%%%%%%%%%%%%%
%%%%%%%%%%%%%%%

\subsubsection{The general case ${\bf \hat{Y}_\nu =i D\, U^\dagger_0}$}
\label{Subsec_ExtendedYnu}

Now, we consider a class of constraints on $Y_\nu$
weaker than the previous one but still very interesting for
models based on flavour symmetries.
We simply assume that at high energy $m_\nu$ can exactly be diagonalized by $U_0$
and $U_0$ is a mass-independent mixing pattern.
The TB mixing pattern, independently from what is the underling flavour symmetry,
is one of the examples of this class of models.
Other examples are given by flavour symmetries which
give rise, at leading order, to the Bi-maximal mixing pattern \cite{BM}, to the golden ratio mixing \cite{golden_ratio} and some (but not all) cases of the trimaximal mixing \cite{Trimaximal}.

In the base where the RH Majorana mass $M_R$ is diagonal
eq.~(\ref{Diag_nu}) can be written as
\beq
\hat m_\nu = -\frac{v^2}{2} U^T_0 \hat{Y}_\nu^T\, \hat M_R^{-1} \,\hat{Y}_\nu U_0~.
\eeq
Defining $X= -i \hat{Y}_\nu U_0$ and supposing that
the masses are not degenerate, the solution to the previous equation is given by
\cite{Lin_lepto,ABMMM_Lepto}
\beq
X =
{\rm diag} (\pm \sqrt{2m_1M_1},\pm \sqrt{2m_2M_2},\pm \sqrt{2m_3M_3})/v~.
\eeq
From this we immediately obtain the constraint (\ref{General_hatY}).
Observe that $\hat{Y}_\nu$ becomes unitary if $D = \unity$. However, the present case is not
strictly a generalization of the previous one since a unitary $Y_\nu$ does not necessarily
imply a mass-independent mixing pattern.

For energy larger than $M_L$ the neutrino mass matrix is fully given by
seesaw formula (\ref{effnumass2}). The initial condition for $m_\nu$
is given by
\beq
m_{\nu (\Lambda)} = U_0^* \hat{m}_\nu U_0^\dagger\;.
\label{EqMnuTBM}
\eeq
In the hatted basis $J_\nu$ is proportional to
\beq
\hat{Y}^\dagger_\nu \hat{Y}_\nu = U_0 D^2 U^\dag_0\;.
\eeq
Since $U_0$ is a mass-independent mixing matrix,
we should expect that the effect of $J_\nu$ is only to
change slightly the mass eigenvalues $m_i$ but not the mixing angles.
In fact, the running effect from $J_\nu$ is then proportional to
\beq
(\hat{Y}^\dagger_\nu \hat{Y}_\nu)^T m_{\nu (\Lambda)} +
{\rm symmetrization} =
\dfrac{2}{v^2}\, U_0^* {\rm diag} (m^2_1M_1, m^2_2M_2, m^2_3M_3)  U_0^\dagger
\eeq
which has exactly the same flavour structure of $m_{\nu (\Lambda)}$.

Now we can move to the energy range between $M_L$ and $M_S$
in which the seesaw formula is only partial as given in eq.~(\ref{effnumass2}).
We can exactly proceed in the same way as the previous case considering
first the running effect from $J_\nu$ on $\accentset{(n)}{\kappa}$ in the hatted basis:
\beq
( \accentset{(n)}{Y}_\nu^\dagger \accentset{(n)}{Y}_\nu )^T\, \accentset{(n)}{\kappa}
~+~{\rm symmetrization}~~~. \nn
\eeq
Explicitly we have
\begin{equation}
2\sum_{m \ne l} \sum_k (U^*_0)_{im}
D^2_m ( U_0^T )_{mk}
(U_0^*)_{kl}\,m_l\,(U^\dagger_0)_{lj} =0 \nn
\end{equation}
where we have used the unitary condition for $U_0$. As a result, this contribution is only global.

Now we move to analyze the second term in (\ref{effnumass2}). Observing that
in the hatted basis we can write
\beq
\accentset{(n)}{Y}_\nu^T\accentset{(n)}M_R^{-1}\accentset{(n)}{Y}_\nu
=  \sum_{m \ne n} (U^*_0)_{im}
(\hat{m})_m (U^\dagger)_{mj} ~~~, \nn
\eeq
using the unitary condition for $U_0$, we obtain
\beq
\sum_{m,p \ne n} \sum_k  (U^*_0)_{ip}
D^2_p ( U_0^T )_{pk}(U^*_0)_{km}
(\hat{m})_m (U^\dagger)_{mj}
= \sum_{m \ne n} (U^*_0)_{km} D^2_m
(\hat{m})_m (U^\dagger)_{mj} ~~~, \nn
\eeq
and the same for the symmetrization part. As we can see, this term also is a global contribution.
Similarly as in the energy range higher than $M_L$, also here, the form
of $m_\nu$ remains invariant and only some of $m_i$ are slightly shifted.
These shifts can be resorbed by redefinition of $m_i$ and do not change anyway the mixing angles
which are contained in $U_0$ and independent from mass eigenvalues.

Then we arrive to a very general conclusion, in any flavour symmetries with
a mass-independent mixing pattern, the running effects from $J_\nu$
correct only the neutrino mass eigenvalues but not the mixing angles.
As in the previous class of models the only flavour-dependent RG contribution to $m_\nu$
is encoded in $J_e$.

%%%%%%%%%%%%%%%%%%%%
%%%%%%%%%%%%%%%%%%%%

\subsubsection{A special case ${\bf U_0 = i U_{\rm TB} P^*}$ and ${\bf D \propto {\rm\bf diag} (1,1,-1)}$}
In this section we consider a special case of $\hat{Y}_\nu =i D\, U^\dagger_0$ in which
the expression of $U_0$ is enforced by a flavour symmetry based on $A_4$ group.
The main feature of this model will be presented in the next section together with
a more detailed analysis of the running effects.
Here we need only the constraint on the mixing matrix $U_0 = i U_{\rm TB} P^*$
and the neutrino Yukawa coupling in the hatted basis:
\beq
\hat{Y}_\nu\equiv yPU_{TB}^TO_{23}=y P\left(\begin{array}{ccc}
                                        \sqrt{2/3}& -1/\sqrt{6} & -1/\sqrt{6}\\
                                        1/\sqrt{3} & +1/\sqrt{3}& +1/\sqrt{3} \\
                                        0 & +1/\sqrt{2} & -1/\sqrt{2}\\
                                        \end{array}\right)
\label{Ynu}
\eeq
where $y$ is a positive parameter of order $\mathcal{O}(1)$, $P$ is a diagonal matrix of phases which corresponds to the Majorana phases and can be written as
\beq
P=\diag(e^{i\al_1/2},\,e^{i\al_2/2},\,e^{i\al_3/2})
\label{Pmatrix}
\eeq
and $O_{23}$ is defined as
\beq
O_{23}=\left(
         \begin{array}{ccc}
           1 & 0 & 0 \\
           0 & 0 & 1 \\
           0 & 1 & 0 \\
         \end{array}
       \right)\;. \nn
\eeq
In order to confront (\ref{Ynu}) with the general expression $\hat{Y}_\nu =i D\, U^\dagger_0$ we observe that
\beq
\hat{Y}_\nu = yPU_{TB}^TO_{23} U_{TB} U_{TB}^T = {\rm diag} (y,y,-y) P U^T_{TB}\;. \nn
\eeq
Then we conclude that (\ref{Ynu}) corresponds to the special case in which $D= {\rm diag} (y,y,-y)$. Furthermore, in the $A_4$ model considered in this paper, there is a very simple relation between $m_i$ and $M_i$ given by $m_i= v^2 y^2 /2M_i$.

Now we explicitly calculate the RG running
from $\La_f$ down to $\lambda$ for this special case
using the approximate analytical expressions given in section \ref{Subsec_Analytical_RGE}.
In the physical basis, the light neutrino mass matrix from eq.~(\ref{EqSeeSaw}) at the initial energy scale $\La_f$ can be recovered by imposing the condition in eq.~(\ref{EqMnuTBM}):
\beq
\begin{split}
m_\nu^{TB}&=-U_{TB}\,P\,\hat{m}_\nu\,P\,U_{TB}^T\\[3mm]
&=-\left[\dfrac{\tilde m_3}{2}\left(\begin{array}{ccc}
                        0&0&0\\
                        0&1&-1\\
                        0&-1&1\end{array}\right)
    +\dfrac{\tilde m_2}{3}\left(\begin{array}{ccc}
                        1&1&1\\
                        1&1&1\\
                        1&1&1\end{array}\right)
    +\dfrac{\tilde m_1}{6}\left(\begin{array}{ccc}
                        4&-2&-2\\
                        -2&1&1\\
                        -2&1&1\end{array}\right)\right]\;,
\end{split}
\label{EqMnuTBMmasses}
\eeq
where $\tilde m_i=m_ie^{i\al_i}$. It is obvious now the meaning of the matrix $P$ as the matrix of the Majorana phases of the light neutrinos.
It is necessary to specify the kind of neutrino mass spectrum: in the Normal Hierarchy (NH) case the light neutrinos are ordered as $m_1<m_2<m_3$ and the heavy ones as $M_3<M_2<M_1$; while in the Inverse Hierarchy (IH) case they are arranged as $m_3<m_1\lesssim m_2$ and $M_2\lesssim M_1<M_3$.

The general result of the running effects on $m_\nu$
is given by eq.~(\ref{generalsol2}) which in our case becomes
\beq
m_{\nu(\lambda)}=I_U(m_\nu^{TB}+\Delta m^{(J_e)}_\nu + \Delta m^{(J_\nu)}_\nu )\;.
\label{mTBM}
\eeq
The analytical result for both $I_U$ and $\Delta m^{(J_e)}_\nu$ (see section \ref{Subsec_Analytical_RGE}) does not depend on the type of the neutrino spectrum,
it is sufficient to identify $M_S,M_M,M_L$ with the correct hierarchy between $M_1, M_2, M_3$~.
In particular, for the TB mixing pattern, the contribution from $J_e$ is given by
\beq
\begin{split}
\Delta m^{(J_e)}_\nu &=m_\nu^{TB} ~ {\rm diag} (0,\, 0,\, \Delta_\tau)+
{\rm diag} (0,\, 0,\, \Delta_\tau) m_\nu^{TB}\\
&=-\left(\begin{array}{ccc}
     0 & 0 & \dfrac{\tilde m_1}{3}-\dfrac{\tilde m_2}{3} \\[2mm]
     0 & 0 & -\dfrac{\tilde m_1}{6}-\dfrac{\tilde m_2}{3}+\dfrac{\tilde m_3}{2}\\[2mm]
     \dfrac{\tilde m_1}{3}-\dfrac{\tilde m_2}{3} & -\dfrac{\tilde m_1}{6}-\dfrac{\tilde m_2}{3}+\dfrac{\tilde m_3}{2} & -\dfrac{\tilde m_1}{3}-\dfrac{2\tilde m_2}{3}-\tilde m_3 \\
         \end{array}
    \right)\Delta_\tau\;.
\end{split}
\eeq
Naturally, the contribution from $J_\nu$ depends on the type of the neutrino spectrum,
however it can be written in the same form for both the spectra:
\beq
\Delta m^{(J_\nu)}_\nu=-\left[\dfrac{\tilde m'_1}{6}\left(
                                                  \begin{array}{ccc}
                                                    4 & -2 & -2 \\
                                                    -2 & 1 & 1 \\
                                                    -2 & 1 & 1 \\
                                                  \end{array}
                                                    \right)
                                +\dfrac{2 \tilde m'_2}{3}\left(
                                                  \begin{array}{ccc}
                                                    1 & 1 & 1 \\
                                                    1 & 1 & 1 \\
                                                    1 & 1 & 1 \\
                                                  \end{array}
                                                    \right)
                                +\tilde m'_3\left(
                                                  \begin{array}{ccc}
                                                    0 & 0 & 0 \\
                                                    0 & 1 & -1 \\
                                                    0 & -1 & 1 \\
                                                  \end{array}
                                                \right)\right]
                                                \label{Deltam_Jnu}
\eeq
where $\tilde m'_i$ are redefinitions of the light neutrino masses:
\begin{description}
\item[\textbf{NH case:}]
\beq
\ba{llll}
\tilde m'_1=\tilde m_1(p+q)\;,& \tilde m'_2=\tilde m_2(x+q)\;,& \tilde m'_3=\tilde m_3(x+z)&\text{in the SM}\\[3mm]
\tilde m'_1=0\;,& \tilde m'_2=2 \tilde m_2x\;,& \tilde m'_3=2 \tilde m_3(x+z)&\text{in the MSSM}
\ea
\label{mpNH}
\eeq
with
\beq
\ba{rcl}
p&=&-\dfrac{1}{16\pi^2}(-3g_2^2+\la+\dfrac{9}{10}g_1^2+\dfrac{9}{2}g_2^2)\ln\dfrac{M_1}{M_2}\\[3mm]
q&=&-\dfrac{1}{16\pi^2}(-3g_2^2+\la+\dfrac{9}{10}g_1^2+\dfrac{9}{2}g_2^2)\ln\dfrac{M_2}{M_3}\\[3mm]
x&=&-\dfrac{y^2}{32\pi^2}\ln\dfrac{M_1}{M_2}\\[3mm]
z&=&-\dfrac{y^2}{32\pi^2}\ln\dfrac{M_2}{M_3}\;;
\ea
\eeq
\item[\textbf{IH case:}]
\beq
\ba{llll}
\tilde m'_1=\tilde m_1(x+q)\;,& \tilde m'_2=\tilde m_2 (x+z)\;,& \tilde m'_3=\tilde m_3(p+q)&\text{in the SM}\\[3mm]
\tilde m'_1=2\tilde m_1 x\;,& \tilde m'_2= 2\tilde m_2 (x+z)\;,& \tilde m'_3=0&\text{in the MSSM}
\ea
\label{mpIH}
\eeq
with
\beq
\ba{rcl}
p&=&-\dfrac{1}{16\pi^2}(-3g_2^2+\la+\dfrac{9}{10}g_1^2+\dfrac{9}{2}g_2^2)\ln\dfrac{M_3}{M_1}\\[3mm]
q&=&-\dfrac{1}{16\pi^2}(-3g_2^2+\la+\dfrac{9}{10}g_1^2+\dfrac{9}{2}g_2^2)\ln\dfrac{M_1}{M_2}\\[3mm]
x&=&-\dfrac{y^2}{32\pi^2}\ln\dfrac{M_3}{M_1}\\[3mm]
z&=&-\dfrac{y^2}{32\pi^2}\ln\dfrac{M_1}{M_2}\;.
\ea
\eeq
\end{description}

Comparing $m_\nu^{TB}$ of eq. (\ref{EqMnuTBMmasses}) with the perturbations $\Delta m_\nu$ of eqs. (\ref{Deltam_Jnu}), we note the presence of the same flavour structure for several matrices and in particular, by redefining $\tilde m_i$ to absorb the terms $\tilde m'_i$ it is possible to account for the seesaw contributions from the RG running into $m_\nu^{TB}$. As a consequence the LO predictions for the TB angles receive corrections only from the terms proportional to $\Delta_\tau$. This result explicitly confirms what we outlined in the previous section.

%
%%%%%%%%%%%%%%%%%%%%%%%%%%%%%%%%%%%%%%%%%%%%%%%       3.2 RGE effects in the charged lepton sector       %%%%%%%%%%%%%%%%%%%%%%%%%%%%%%%%%%%
%

\subsection{RGE effects in the charged lepton sector}
\label{Subsec_Chargedsector}

The presence of a term proportional to $\hat{Y}^\dagger_\nu \hat{Y}_\nu$ in the RG equation
for $Y_e$ can switch on off-diagonal entries in the charged lepton Yukawa matrix $Y_e$~.
When rotated away, this additional contribution introduces a non-trivial $U_e$
and consequently corrects the lepton mixing matrix $U_{PMNS}$.
For a unitary $\hat Y_\nu$, this correction appears only between the seesaw mass scales
while, in the general case discussed in section \ref{Subsec_ExtendedYnu}~, it appears
already from the cutoff $\Lambda_f$~.

In close analogy with the running effects on neutrino mass matrix (\ref{mTBM}),
the full result of the running for charged lepton mass matrix can conventionally
be written as
\beq
(Y^\dagger_e Y_e)_ {(\lambda)}=I_e\left[ (Y^\dagger_e Y_e)_{(\Lambda_f)}
+\Delta (Y^\dagger_e Y_e) \right]\;,
\label{CorrectedYeYe}
\eeq
where $I_e$ is an irrelevant global coefficient which can be absorbed
by, for example, $y_\tau$~.
Now we move to the case of TB mixing pattern.
In this case, the flavour-dependent corrections can be
explicitly calculated:
\begin{description}
\item[\textbf{NH case:}]
\beq
\Delta (Y^\dagger_e Y_e)\simeq y_\tau^2 \left[ a_e \left(
                                                  \begin{array}{ccc}
                                                    0 & 0 & 1 \\
                                                    0 & 0 & -\frac12 \\
                                                    1 & -\frac12 & 5 \\
                                                  \end{array}
                                                    \right)
                                		+b_e \left(
                                                  \begin{array}{ccc}
                                                    0 & 0 & 0 \\
                                                    0 & 0 & -1 \\
                                                    0 & -1 & 2 \\
                                                  \end{array}
                                                    \right)
                                		+c_e \left(
                                                  \begin{array}{ccc}
                                                    0 & 0 & 0 \\
                                                    0 & 0 & 0 \\
                                                    0 & 0 & 2 \\
                                                  \end{array}
						    \right)\right]\;,	
\label{EqDeltaYeDagYeNH}
\eeq
\item[\textbf{IH case:}]
\beq
\Delta (Y^\dagger_e Y_e)\simeq y_\tau^2 \left[a'_e  \left(
                                                  \begin{array}{ccc}
                                                    0 & 0 & 0 \\
                                                    0 & 0 & 1 \\
                                                    0 & 1 & 2 \\
                                                  \end{array}
                                                    \right)
                                		+b'_e \left(
                                                  \begin{array}{ccc}
                                                    0 & 0 & 1\\
                                                    0 & 0 & 1 \\
                                                    1 & 1 & 2 \\
                                                  \end{array}
                                                    \right)
                                		+c'_e \left(
                                                  \begin{array}{ccc}
                                                    0 & 0 & 0 \\
                                                    0 & 0 & 0 \\
                                                    0 & 0 & 2 \\
                                                  \end{array}
                                                \right)\right]\;,
\label{EqDeltaYeDagYeIH}
\eeq
\end{description}
where the coefficients are
\beq
\ba{lr}
a_e=b'_e=-\dfrac{C'_\nu}{16 \pi^2} \dfrac{y^2}{3} \ln\dfrac{M_1}{M_2},&\qquad b_e=-\dfrac{C'_\nu}{16 \pi^2} \dfrac{y^2}{2} \ln\dfrac{M_2}{M_3}\;, \\[5mm]
c_e=c'_e=-\dfrac{3 C'_e y_\tau^2}{16 \pi^2}\ln\dfrac{\Lambda_f}{m_\mathrm{SUSY}(m_Z)}\;,&\qquad a'_e= -\dfrac{C'_\nu}{16 \pi^2} \dfrac{y^2}{2} \ln\dfrac{M_3}{M_1}\;,
\label{coeff}
\ea
\eeq
and $C'_\nu=-3/2 \;(1)$, $C'_e=3/2 \;(3)$  in the SM (MSSM).
Here we observe that the off-diagonal contributions to $Y^\dagger_e Y_e$
are encoded in $a_e$, $b_e$, $a'_e$ and $b'_e$ which depend only
on the seesaw scales $M_i$~. As a result, as we will show in the
next section, $c_e$ and $c'_e$ do not affect the lepton mixing angles.

%
%%%%%%%%%%%%%%%%%%%%%%%%%%%%%%%%%%%%%%%%%%%%%%%       3.3 Full RGE effects on the TB mixing pattern    %%%%%%%%%%%%%%%%%%%%%%%%%%%%%%%%%%%
%

\subsection{Full RGE effects on the TB mixing pattern}
\label{RGcoefficients}

In this section, we combine various contributions discussed in previous sections into the observable matrix $U_{PMNS}$ from which we extract angles and phases at low energy.
Since we are interested in physical quantities, we eliminate one of the phases of $P$
defined in (\ref{Pmatrix}) and in particular we express each result in function of $\alpha_{ij}\equiv(\alpha_i-\alpha_j)/2$,
removing $\alpha_3$. The corrected mixing angles can be written as
\beq
\theta_{ij(m_\la)} = \theta^{TB}_{ij}+k_{ij}+\ldots
\eeq
where $\theta^{TB}_{13} = 0$,  $\theta^{TB}_{12} = \arcsin \sqrt{1/3}$, $\theta^{TB}_{23} = -\pi/4$ and $k_{ij}$ are defined by
\beq
\begin{split}
k_{12}=&\dfrac{1}{3\sqrt2}\left(\dfrac{|\tilde m_1+\tilde m_2|^2}{m_2^2-m_1^2} \Delta_\tau - 3 a_e\right) \\[5mm]
k_{23}=&\dfrac{1}{6}\left[\left(\dfrac{|\tilde m_1+\tilde m_3|^2}{m_3^2-m_1^2}+2\dfrac{|\tilde m_2+\tilde m_3|^2}{m_3^2-m_2^2}\right) \Delta_\tau -3 a_e -6 b_e \right] \qquad \qquad \text{for NH} \\[5mm]
=&\dfrac{1}{6}\left[\left(\dfrac{|\tilde m_1+\tilde m_3|^2}{m_3^2-m_1^2}+2\dfrac{|\tilde m_2+\tilde m_3|^2}{m_3^2-m_2^2}\right) \Delta_\tau +3 a_e +3 a'_e \right] \qquad \qquad \text{for IH} \\[5mm]
k_{13}=&\dfrac{1}{3\sqrt2}\sqrt{4m_3^2 \Delta_\tau^2\left(\dfrac{m_1\sin\alpha_{13}}{m_1^2-m_3^2}-\dfrac{m_2\sin\alpha_{23}}{m_2^2-m_3^2}\right)^2+ \left[\left(\dfrac{|\tilde m_1+\tilde m_3|^2}{m_1^2-m_3^2}-\dfrac{|\tilde m_2+\tilde m_3|^2}{\tilde m_2^2-\tilde m_3^2}\right)\Delta_\tau-3 a_e\right]^2}\;
\end{split}
\nn
\eeq
and the dots stand for sub-leading corrections.
In the previous expressions we can clearly distinguish the contributions
coming from the diagonalization of the corrected TB neutrino mass matrix
(\ref{mTBM}) and those from the diagonalization of (\ref{CorrectedYeYe}).
As it is clear from (\ref{coeff}), the corrections to the TB mixing from the charged lepton
sector is important only for hierarchical RH neutrinos and will approach to zero
as soon as the spectrum becomes degenerate.
On the other hand, the corrections from the neutrino sector
should be enhanced if the light neutrinos are quasi-degenerate and if the $\tan \beta$ is large, in the MSSM case.

The physical Majorana phases are also corrected due to the RG running and we found the
following results:
\beq
\alpha_{ij(m_\la)}\simeq\alpha_{ij}+\delta\alpha_{ij}\Delta_\tau+\ldots
\eeq
where $\alpha_{ij}$ are the starting values at $\La_f$ and
\bea
\delta\alpha_{13}&=&\dfrac{2}{3}\dfrac{m_1m_2\sin(\alpha_{13}-\alpha_{23})}{m_2^2-m_1^2}
\label{Deltaalpha13} \\[5mm]
\delta\alpha_{23}&=&\dfrac{4}{3}\dfrac{m_1m_2\sin(\alpha_{13}-\alpha_{23})}{m_2^2-m_1^2}
\label{Deltaalpha23} \;.
\eea
At $\La_f$, $\sin{\theta^{TB}_{13}}$ is vanishing and as a result the Dirac CP-violating phase is undetermined. An alternative is to study the Jarlskog invariants \cite{Jarlskog} which are well-defined at each energy scale:
\beq
J_{CP}=\dfrac{1}{2}\left|\Im\left\{(U_{PMNS})^*_{ii}(U_{PMNS})_{ij}(U_{PMNS})_{ji}(U_{PMNS})^*_{jj}\right\}\right|\;,
\eeq
where $i,j\in\{1,2,3\}$ and $i\neq j$. At $\La_f$, $J_{CP}$ is vanishing, while after the RG running it is given by
\beq
J_{CP}=\dfrac{1}{18}\left|m_3\left(\dfrac{m_1\sin\alpha_{13}}{m_1^2-m_3^2}- \dfrac{m_2\sin\alpha_{23}}{m_2^2-m_3^2}\right)\right|\Delta_\tau\;.
\label{Jarlskog}
\eeq
Two comments are worth. First of all, in the expression for $k_{13}$, it is easy to recover the resulting expression for $J_{CP}$ as the first term under the square root, a part global coefficients. This means that the RG procedure introduce a mixing between the expression of the reactor angle and of the Dirac CP-phase. Moreover we can recover the value of the Dirac CP-phase directly from eq. (\ref{Jarlskog}) and we get the following expression:
\beq
\begin{split}
\cot\delta_{CP}=&-\dfrac{m_1(m_2^2-m_3^2)\cos\alpha_{13}-m_2(m_1^2-m_3^2) \cos\alpha_{23}-m_3(m_1^2-m_2^2)}{m_1(m_2^2-m_3^2)\sin\alpha_{13}-m_2(m_1^2-m_3^2)\sin\alpha_{23}}+\\[3mm]
&-\dfrac{3 a_e (m_2^2-m_3^2)(m_1^2-m_3^2)}{2 m_3 \left[m_1(m_2^2-m_3^2)\sin\alpha_{13}-m_2(m_1^2-m_3^2)\sin\alpha_{23}\right] \Delta_\tau}  \;.
\end{split}
\eeq
In the neutrino sector, the RG contributions from the See-Saw terms are present only in the resulting mass eigenvalues:
\beq
m_{i(\la)}\simeq m_i(1+\delta m_i)+\ldots
\eeq
where $m_i$ are the starting values at $\La_f$ and $\delta m_i$, in both the SM and the MSSM and in both the NH and IH spectra, are given by
\beq
\delta m_1=\dfrac{m'_1}{m_1}-\dfrac{\Delta_\tau}{3}\;,\qquad
\delta m_2=2\dfrac{m'_2}{m_2}-\dfrac{2\Delta_\tau}{3}\;,\qquad
\delta m_3=2\dfrac{m'_3}{m_3}-\Delta_\tau\;,
\eeq
with $m'_i\equiv|\tilde m'_i|$, given as in eqs. (\ref{mpNH}, \ref{mpIH}).

%%%%%%%%%%%%%%%%%%%%%%%%%%%%%%%%%%%%%%%%%%%%%%%%%%%%%%%%%%%%%%%%%%%%%%%%%%%%%%%%%%%%%%%%%%%%%%%%%%%%%%%%%%%%%%%%%%%%%%%%%%%%%%%%%%%%%%%%%
%%%%%%%%%%%%%%%%%%%%%%%%%%%%%%%%%%%%%%%%%%%%%%%       3. The Altarelli-Feruglio model       %%%%%%%%%%%%%%%%%%%%%%%%%%%%%%%%%%%%%%%%%%%%%
%%%%%%%%%%%%%%%%%%%%%%%%%%%%%%%%%%%%%%%%%%%%%%%%%%%%%%%%%%%%%%%%%%%%%%%%%%%%%%%%%%%%%%%%%%%%%%%%%%%%%%%%%%%%%%%%%%%%%%%%%%%%%%%%%%%%%%%%%

\section{The Altarelli-Feruglio (AF) model}
\label{Sec_AFmodel}

We recall here the main features of the AF model \cite{AF_Papers}, which is based on the flavour group $G_f=A_4\times Z_3\times U(1)_{FN}$: the spontaneous breaking of $A_4$ is responsible for the TB mixing; the cyclic symmetry $Z_3$ prevents the appearance of dangerous couplings and helps keeping separated the charged lepton sector and the neutrino one; the $U(1)_{FN}$ \cite{FN} provides a natural hierarchy among the charged lepton masses. $A_4$ is the group of the even permutations of $4$ objects and it has been largely studied in literature \cite{A4I,A4II}: it has 12 elements and 4 inequivalent irreducible representations, three singlets $1$, $1'$, $1''$, and a triplet $3$. We refer to \cite{AF_Papers} for group details and recall here only the multiplication rules:
\beq
\begin{array}{ll}
1\otimes R&=R\quad \textrm{with \emph{R} any representation}\\
1'\otimes1'&=1''\\
1''\otimes1''&=1'\\
1'\otimes1''&=1\\
\\
3\otimes3&=1\oplus1'\oplus1''\oplus3_S\oplus3_A\;.
\end{array}
\eeq
The TB mixing is achieved through a well-defined symmetry breaking mechanism: $A_4$ is spontaneously broken down to $G_\nu=Z_2$ in the neutrino sector and to a different subgroup $G_\ell=Z_3$ in the charged lepton one. This breaking chain is fundamental in the model, because $G_\nu$ and $G_\ell$ represent the low-energy flavour structures of neutrinos and charged leptons, respectively. This mechanism is produced by a set of scalar fields, the flavons, which transform only under the flavour group $G_f$. In table \ref{table:transformations}, we can see the fermion and the scalar content of the model and their transformation properties under $G_f$.

\begin{table}[ht]
\begin{center}
\begin{tabular}{|c||ccccc||ccccc|}
\hline
&&&&&&&&&&\\[-4mm]
 & $\ell$ & $e^c$ & $\mu^c$ & $\tau^c$ & $\nu^c$ &$H$ & $\theta$ & $\phit$ & $\phis$ & $\xi$ \\[2mm]
\hline
&&&&&&&&&&\\[-4mm]
$A_4$ & 3 & 1 & $\onepp$ & $\onep$ & 3 & 1 & 1 & 3 & 3 & 1 \\[2mm]
$Z_3$ & $\om$ & $\om^2$ & $\om^2$ & $\om^2$ & $\om^2$ & 1 & 1 & 1 & $\om^2$ & $\om^2$ \\[2mm]
$U(1)_{FN}$ & 0 & 2 & 1 & 0 & 0 & 0 & -1 & 0 & 0 & 0  \\[2mm]
\hline
\end{tabular}
\end{center}
\caption{\label{table:transformations} The transformation properties of the fields under $A_4$, $Z_3$ and $U(1)_{FN}$.}
\end{table}

The specific breaking patter of the symmetry which leads to the TB scheme for leptons requires that the flavons develop VEVs with a specific alignment:
\beq
\ba{ccl}
\dfrac{\langle\phit\rangle}{\Lambda_f}&=&(u,0,0)\\[3mm]
\dfrac{\langle\phis\rangle}{\Lambda_f}&=&c_b(u,u,u)\\[3mm]
\dfrac{\langle\xi\rangle}{\Lambda_f}&=&c_a u\\[3mm]
\dfrac{\langle\theta_{FN}\rangle}{\Lambda_f}&=&t
\ea
\label{vevs}
\eeq
where $c_{a,b}$ are complex numbers with absolute value of order one, while $u$ and $t$ are the small symmetry breaking parameters of the theory  which parametrise the ratio of the VEVs of the flavons over the cutoff $\Lambda_f$ of the theory (they can be taken real through field redefinitions). In \cite{AF_Papers} it has been shown a natural explanation of this misalignment.

Once defined the transformations of all the fields under $G_f$, it is possible to write down the Yukawa interactions: at the LO they read
\bea
{\cal L}_\ell^{LO}&=&\dfrac{y_e}{\Lambda^3} \theta^2e^c H^\dagger \left(\varphi_T \ell\right)
+\dfrac{y_\mu}{\Lambda^2} \theta\mu^c H^\dagger \left(\varphi_T \ell\right)'
+\dfrac{y_\tau}{\Lambda} \tau^c H^\dagger \left(\varphi_T \ell\right)''+h.c.
\label{Ll}\\
\nn\\
{\cal L}_\nu^{LO}&=& y(\nu^{c\,T} \widetilde H^\dag \ell)+x_a\xi(\nu^{c\,T}\nu^c)+x_b(\varphi_S\nu^{c\,T}\nu^c)+h.c.\;.
\label{Lnu}
\eea
The notation $(\ldots)$, $(\ldots)'$ and $(\ldots)''$ refers to the contractions in $1$, $1'$ and $1''$, respectively. When the flavons develop VEVs in agreement with eq. (\ref{vevs}) and after the electroweak symmetry breaking, the LO mass matrix of charged leptons takes the following form:
\beq
m_\ell=\dfrac{v\, u}{\sqrt2}\left(
\begin{array}{ccc}
y_e t^2 & 0& 0\\
0& y_\mu t& 0\\
0& 0& y_\tau
\end{array}
\right) \;,
\label{yf}
\eeq
with $y_e$, $y_\mu$ and $y_\tau$ being complex numbers with absolute values of order one. As one can see the relative hierarchy among the charged lepton masses is given by the parameter $t$: when
\beq
t\approx 0.05
\label{tbound}
\eeq
then the mass hierarchy is in agreement with the experimental measurements. As we will see in the following sections, the model admits a well defined range for the parameter $u$ which can approximatively be set to
\beq
0.003\lesssim u \lesssim 0.05\;.
\eeq

In the neutrino sector, the Dirac and the Majorana mass matrices, at the LO, are given by
\beq
m_D\equiv \dfrac{y\,v}{\sqrt2} \left(
              \begin{array}{ccc}
                1 & 0 & 0 \\
                0 & 0 & 1 \\
                0 & 1 & 0 \\
              \end{array}
            \right)\qquad
M_R \propto \left(
\begin{array}{ccc}
a+2 b/3& -b/3& -b/3\\
-b/3& 2b/3& a-b/3\\
-b/3& a-b/3& 2 b/3
\end{array}
\right)\;,
\label{Y}
\eeq
where $a\equiv 2x_a c_a u$ and $b\equiv 2x_b c_b u$. The complex symmetric matrix $M_R$ is diagonalised by the transformation
\beq
\hat{M}_R=U_R^T M_R U_R\;,
\label{Mhat}
\eeq
where $\hat M_R$ is a diagonal matrix with real and positive entries, given by
\beq
\hat M_R\equiv \diag(M_1,\,M_2,\,M_3)=\diag(|a+b|,|a|,|-a+b|)\;,
\label{RHEigenvalues}
\eeq
while the unitary matrix $U_R$ can be written as,
\beq
U_R=U_{TB}P\;,
\eeq
where P is the diagonal matrix of the Majorana phases already defined in eq. (\ref{Pmatrix}), with $\al_1=-\arg(a+b)$, $\al_2=-arg(a)$, $\al_3=-\arg(-a+b)$.
After the electroweak symmetry breaking, the mass matrix for the light neutrinos is recovered from the well known type I seesaw formula
\beq
m_\nu=-m_D^T M_R^{-1}m_D=-\dfrac{v^2\,y^2}{2}M_R^{-1}\;
\label{mnu}
\eeq
where the last passage is possible considering that $M_R^{-1} m_D=m_D M_R^{-1}$. From \eq{Mhat}, $U_R^\dag M_R^{-1} U_R^*=\diag(M_1^{-1},M_2^{-1},M_3^{-1})$ and as a result the light neutrino mass matrix can be diagonalised by
\beq
\hat{m}_\nu=U_\nu^Tm_\nu U_\nu\;,
\eeq
where $U_\nu=iU_R^*=iU_{TB}P^*$ and the diagonal matrix $\hat{m}_\nu$ has real and positive entries written as \footnote{Note that we can absorb the phase of $y$ in $P$ without loss of generality: it is a global non-observable phase.}
\beq
m_i=\dfrac{v^2}{2}\dfrac{y^2}{M_i}\;.
\label{LightMasses}
\eeq

When the SUSY context is considered $G_f$ accounts for an additional term, a continuous $R$ symmetry $U(1)_R$ which simplifies the constructions of the scalar potential: under this symmetry, the matter superfields are single-valued, while the scalar ones are neutral.
It is easy to extend eqs. (\ref{Ll},\ref{Lnu}) in the supersymmetric case: two Higgs doublets $h_{d,u}$, invariant under $A_4$, substitute $H$ and $\widetilde H$, respectively; the Lagrangian $\cL_\ell^{LO}$ is identified to the LO charge lepton superpotential $w_\ell^{LO}$ and $\cL_\nu^{LO}$ is identified to the LO neutrino superpotential $w_\nu^{LO}$.
While $t$ is still equal to $0.05$ in order to have a correct charged lepton mass hierarchy, the range for $u$ slightly changes:
\beq
0.007\lesssim u \lesssim 0.05\;.
\eeq

%%%%%%%%%%%%%%%%%%%%%%%%%%%%%%%%%%%%%%%%%%%%%%%       4.1 The light neutrino mass spectrum            %%%%%%%%%%%%%%%%%%%%%

\subsection{The light neutrino mass spectrum}

Following \cite{BBFN_Lepto,BMM_SS}, we summarize the results for the light neutrino mass spectrum, which will be useful to discuss the effects of the running on the low-energy observables. We note that the following analysis is valid in the SM as well as in the MSSM. The light neutrino masses are directly linked to the heavy neutrino masses through \eq{LightMasses} and they can be expressed in terms of only three independent parameters. It is possible to choose these parameters in order to simply the analysis: $|a|=|M_2|=v^2|y|^2/(2|m_2|)$ (in the MSSM we replace $v$ with $v_u$ ), $\rho$ and $\Delta$, where $\rho$ and $\Delta$ are defined as
\beq
\dfrac{b}{a}=\rho\,e^{i\Delta}\;,
\eeq
where $\Delta$ is defined in the range $[0,\,2\pi]$.
From the experimental side only the squared mass differences have been measured: for the NH (IH) they are \cite{Data}
\beq
\ba{rcl}
\Delta m^2_\mathrm{sol}&\equiv& m_2^2-m_1^2\\[3mm]
\Delta m^2_\mathrm{atm}&\equiv& |m_3^2-m_1^2 (m_2^2)|\;.
\ea
\label{DeltaNuMasses}
\eeq
As a result the spectrum is not fully determined and indeed $\Delta$ is still a free parameter. We can bound this parameter, requiring $|\cos\Delta|\leq1$. In order to get analytical relations for $\rho$ and $\cos\Delta$, we calculate the following mass ratios:
\baq
\dfrac{m_2^2}{m_{1(3)}^2}=1\pm2 \rho\cos\Delta+\rho^2\;.
\eaq
It is then easy to express $\rho$ and $\cos\Delta$ as a function of the neutrino masses:
\beq
\rho=\sqrt{\dfrac{1}{2}\left(\dfrac{m_2^2}{m_1^2}+\dfrac{m_2^2}{m_3^2}\right)-1}\qquad\qquad
\cos\Delta=\dfrac{\dfrac{m_2^2}{m_1^2}-\dfrac{m_2^2}{m_1^2}}{4\sqrt{\dfrac{1}{2}\left(\dfrac{m_2^2}{m_1^2}+\dfrac{m_2^2}{m_3^2}\right)-1}}\;.
\eeq
It is interesting to note that this expression holds for both the types of spectra. Using eqs. (\ref{DeltaNuMasses}), it is possible to express $\cos\Delta$ in function of only the lightest neutrino mass and, imposing the constraint $|\cos\Delta|\leq1$, the following ranges can be derived: taking the central values of $\Delta m^2_\mathrm{sol}$ and $\Delta m^2_\mathrm{atm}$
\baq
\mathrm{NH:}\qquad&4.46\;\mathrm{meV}<m_1<5.91\;\mathrm{meV}&\\[3mm]
\mathrm{IH:}\qquad&17.1\;\mathrm{meV}<m_3\;.\phantom{<5.91\mathrm{meV}}&
\label{RangeMasses}
\eaq
For the NH, $m_1$ spans in a narrow range of values, which corresponds to values of $\Delta$ close to zero. On the other hand, for the IH, $m_3$ is bounded only from below and the minimum is achieved when $\Delta$ is close to $\pm\pi$. Furthermore, $\cos\Delta$ is restricted only to negative values (see \cite{BMM_SS,BBFN_Lepto} for further details).

\begin{figure}[ht!]
 \centering
\includegraphics[width=7.7cm]{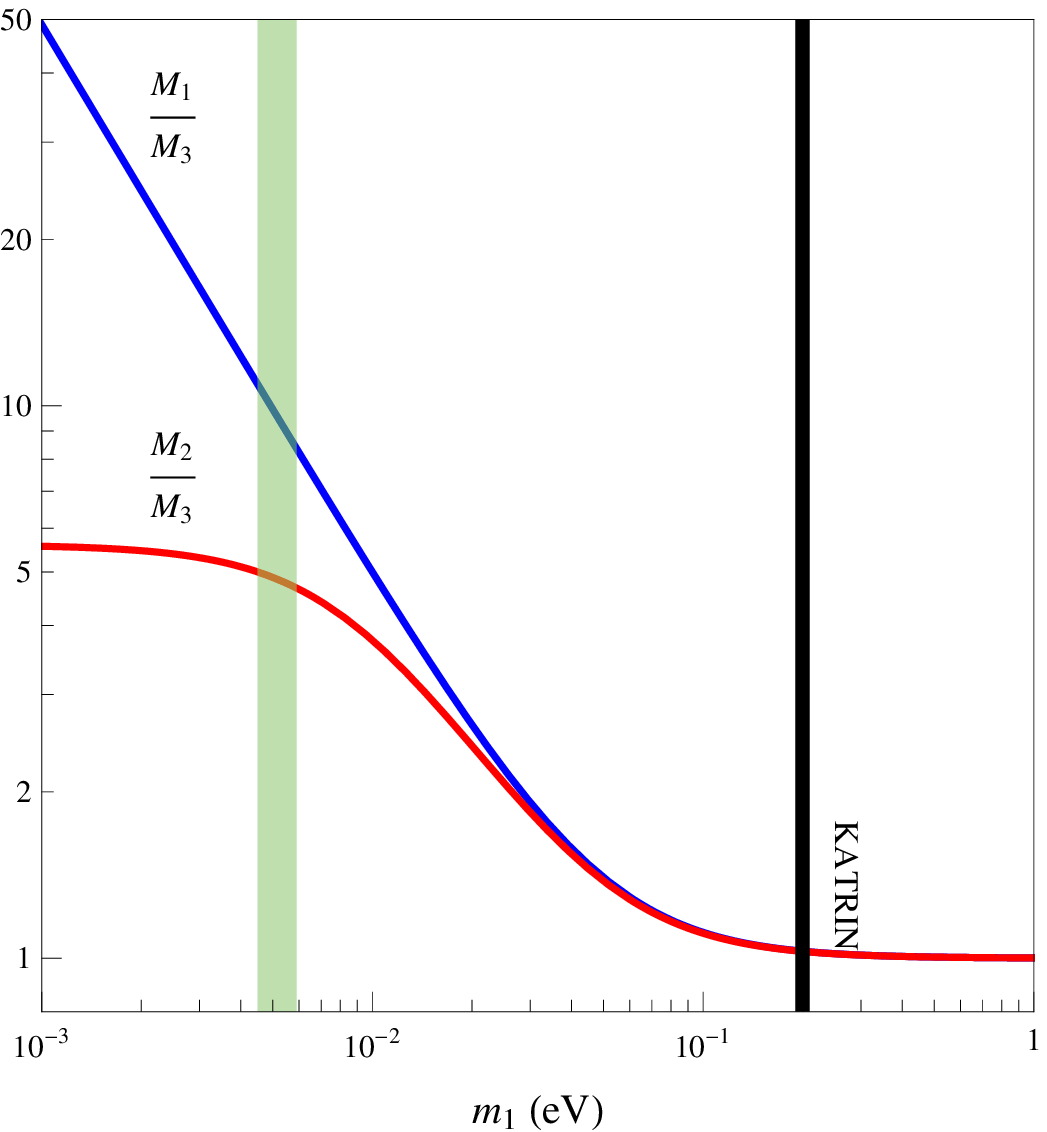}
\includegraphics[width=7.7cm]{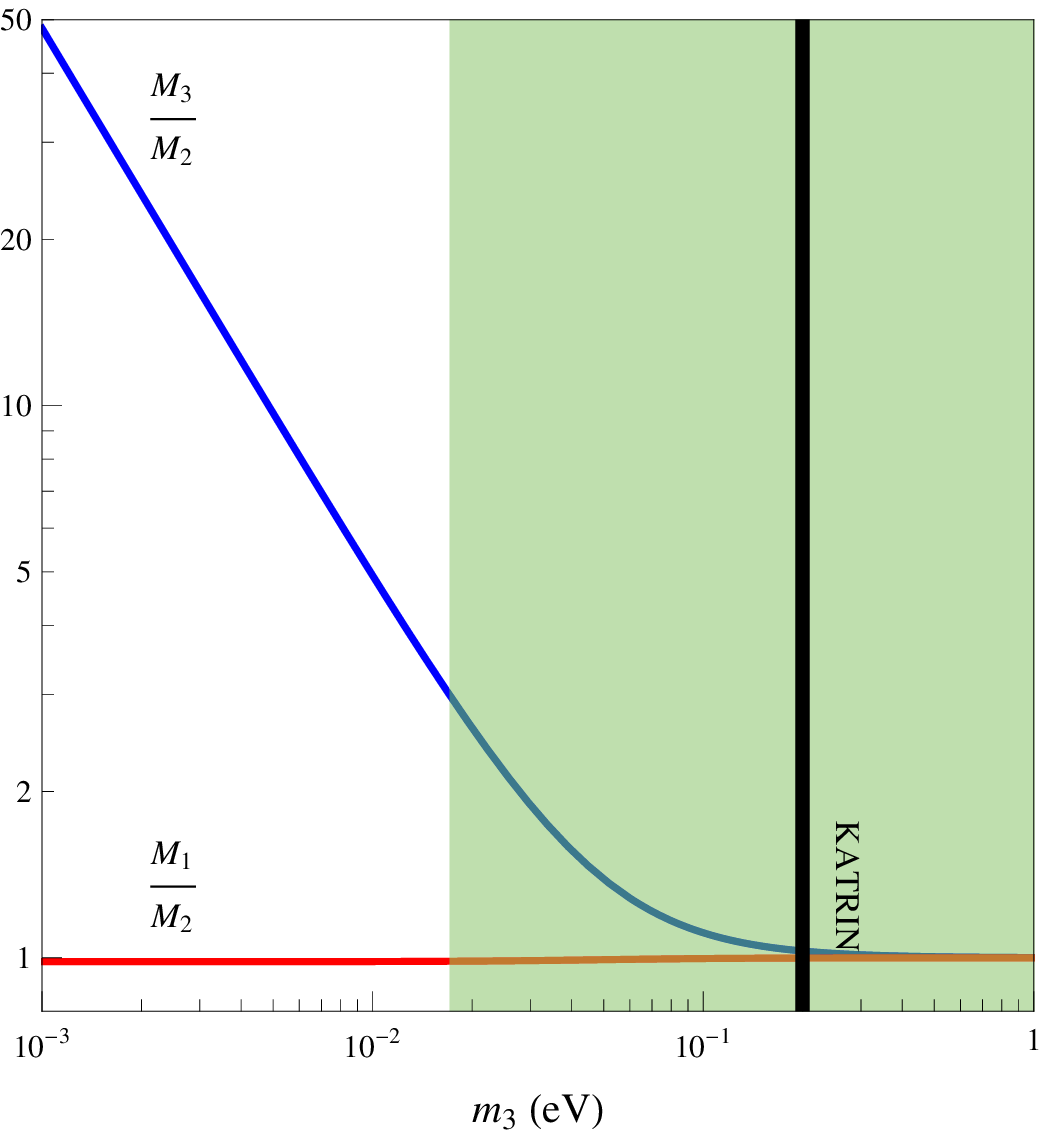}
\caption{Plots of the ratios of the heavy RH neutrino masses as a function of the lightest LH neutrino masses. On the left the NH case and on the right the IH one. The green opaque areas refer to the allowed range for $m_{1(3)}$ as in eq. (\ref{RangeMasses}). The vertical black lines correspond to the future sensitivity of $0.2$ eV of KATRIN experiment.}
 \label{fig:RH}
\end{figure}

From eq. (\ref{LightMasses}) it is possible to describe the LO spectrum of the RH neutrinos in function of a unique single parameter, which is the lightest LH neutrino mass. In all the allowed range for $m_{1(3)}$, the order of magnitude of the RH neutrino masses is $10^{14\div15}$ GeV. In fig. (\ref{fig:RH}) we show explicitly the ratios of the RH neutrino masses for NH and the IH cases, on the left and on the right respectively. The ratios are well defined for the NH, thanks to the narrow allowed range for $m_1$: $M_1/M_3\sim11$ and $M_2/M_3\sim5$. In the case of the IH, the ratio $M_1/M_2$ is fixed at $1$ while $M_3/M_2$ varies from about $3$ to $1$, going from the lower bound of $m_3$ up to the KATRIN sensitivity.

These results are valid only at LO and some deviations are expected with the introduction of the higher order terms, that is illustrated in the following section. The result of a direct computation shows that for the NH spectrum the corrections leave approximatively unaffected eqs. (\ref{RangeMasses}); this is true for the IH case too, apart when the neutrino masses reach values at about $0.1$ eV for which the deviations become significant.\\

We only commented on the neutrino masses, both light and heavy, and we saw that it is possible to express all of them in function of the lightest LH neutrino mass. The same is true for the phases too: since in the TB mixing the reactor angle is vanishing, the Dirac CP phase is undetermined at LO; on the contrary the Majorana phases are well defined and they can be expressed through $\rho$ and $\Delta$. Since we are interested in physical observables, we report only phases differences, $\alpha_{ij}\equiv(\alpha_i-\alpha_j)/2$:
\baq
\sin(2\alpha_{23})&=&\dfrac{\rho\sin\Delta}{\sqrt{1-2\rho\cos\Delta+\rho^2}}\\[5mm]
\sin(2\alpha_{13})&=&\dfrac{2\rho\sin\Delta}{\sqrt{(\rho^2-1)^2+4\rho^2\sin^2\Delta}}\;.
\label{MajoranaPhases}
\eaq
It will be useful for the following discussion to show also $\sin(2\alpha_{12})$, which enters in the RG evolution of the physical Majorana phases:
\beq
\sin(2\alpha_{12})=-\dfrac{\rho\sin\Delta}{\sqrt{1+2\rho\cos\Delta+\rho^2}}\;.
\eeq

%%%%%%%%%%%%%%%%%%%%%%%%%%%%%%%%%%%%%%%%%%%%%%%       4.2 The next-to-leading order contributions       %%%%%%%%%%%%%%%%%%%%%%%%%%%%%%%%%%

\subsection{The next-to-leading order (NLO) contributions}
\label{NLO}

Another important implication of the spontaneously broken flavour symmetry is that the leading order predictions are always subjected to corrections due to higher-dimensional operators. The latter are suppressed by additional powers of the cutoff $\Lambda_f$ and can be organized in a suitable double power expansion in $u$ and $t$.

At the NLO there are may additional terms which can be added in the Lagrangian.
Since $\varphi_T$ is the only scalar field which
is neutral under the abelian part of the flavour symmetry,
all the NLO terms contain the terms already present in the LO Lagrangian
multiplied by an additional $\varphi_T /\Lambda_f$~.
In addition to these terms, there are also corrections to the leading vacuum alignment
in eq. (\ref{vevs}):
\beq
\ba{ccl}
\dfrac{\langle\phit\rangle}{\Lambda_f}&=&(u,0,0)+(c_1 u^2,c_2 u^2,c_3 u^2)\\[3mm]
\dfrac{\langle\phis\rangle}{\Lambda_f}&=&c_b(u,u,u)+(c_4 u^2,c_5 u^2,c_6 u^2)\\[3mm]
\dfrac{\langle\xi\rangle}{\Lambda_f}&=&c_a u+c_7 u^2
\ea
\label{vevsplus}
\eeq
where $c_i$ are complex numbers with absolute value of order one.
Note that in the MSSM, the model predicts $c_2=c_3$.
Here we will not perform a detailed analysis for NLO operators and
the origin of (\ref{vevsplus}), see  \cite{AF_Papers, BBFN_Lepto} for a relative study.
As a result, all the quantities relevant in this paper receive a
NLO correction at the $\mathcal{O}(u)$ level:
\beq
Y_e + \delta Y_e\;, \qquad Y_\nu + \delta Y_\nu \; , \qquad M + \delta M \nn\;.
\eeq
These corrections should modify the resulting $m_\nu$ and the mixing angles
in addition to the RG contributions.
It is not difficult to see that deviations from TB induced only by NLO corrections
are also of the same level \cite{AF_Papers, BBFN_Lepto}:
\beq
\sin ^2 \theta_{23} = \frac 12 + \mathcal{O}(u), \qquad \sin^2 \theta_{12}=\frac 13 + \mathcal{O}(u),
\qquad \sin \theta_{13} = \mathcal{O}(u).
\label{Uhighcorr}
\eeq
Since the solar mixing angle is, at present, the most precisely known, we require that its value remains at $3\sigma$ range \cite{Data}.
This requirement results in an upper bound on $u$ of about $0.05$.
On the other hand we have from eq. (\ref{yf}) the following relation:
\beq
\ba{rll}
u&=\, \dfrac{1}{|y_\tau|} \dd\frac{\sqrt{2} m_\tau}{v}\approx 0.01 \dd\frac{1}{|y_\tau|}&\qquad\text{in the SM}\\[5mm]
u&\simeq\,\dfrac{\tan\beta}{|y_\tau|} \dd\frac{\sqrt{2} m_\tau}{v} \approx 0.01 \dd\frac{\tan\beta}{|y_\tau|}&\qquad\text{in the MSSM}
\label{tanb&u&yt}
\ea
\eeq
where for the $\tau$ lepton we have used its pole mass $m_\tau=(1776.84 \pm 0.17) \;\rm{MeV}$ \cite{pdg2008}. Requesting $|y_\tau|<3$ we find a lower limit for $u$ of about $0.003$ in the SM case; in the MSSM, the same requirements provides a lower bound close to the upper bound $0.05$ for $\tan\beta=15$, whereas for $\tan\beta=2$ it is $u>0.007$. From now on, we will choose the maximal range of $u$ as
\beq
0.003< u < 0.05
\label{uboundSM}
\eeq
for the SM context, while for the MSSM one we take
\beq
0.007< u < 0.05\;,
\label{ubound}
\eeq
which shrinks when $\tan\beta$ is increased from 2 to 15.

The NLO terms affect also the previous results for the Dirac and the Majorana phases. All the new parameters which perturb the LO results are complex and therefore they introduce corrections to the phases of the PMNS matrix. Due to the large amount of such a parameters, we expect large deviations from the LO values.

%%%%%%%%%%%%%%%%%%%%%%%%%%%%%%%%%%%%%%%%%%%%%%%%%%%%%%%%%%%%%%%%%%%%%%%%%%%%%%%%%%%%%%%%%%%%%%%%%%%%%%%%%%%%%%%%%%%%%%%%%%%%%%%%%%%%%%%%%
%%%%%%%%%%%%%%%%%%%%%%%%%%%%%%%%%%%%%%%%%%%%%%%       5. RG effects on mixing angles in the AF model   %%%%%%%%%%%%%%%%%%%%%%%%%%%%%%%%%%
%%%%%%%%%%%%%%%%%%%%%%%%%%%%%%%%%%%%%%%%%%%%%%%%%%%%%%%%%%%%%%%%%%%%%%%%%%%%%%%%%%%%%%%%%%%%%%%%%%%%%%%%%%%%%%%%%%%%%%%%%%%%%%%%%%%%%%%%%

\section{RG effects in the Altarelli-Feruglio model}
\label{Sec_AFmodel_RG}

In this section we will apply the analysis of RG running effects on the lepton mixing angles
to the AF model. In order to perform such a study, it is important to verify the initial assumptions made in section \ref{RGcoefficients}, in particular, we see that eq.~(\ref{Ynu}) exactly corresponds to the one implied by the AF model, when moving to the physical basis.
On the other side, the presence of flavon fields has a relevant impact on the results of the analysis.
In the unbroken phase, flavons are active fields and should modify the RG equations. Although the study of the relevant Feynman diagrams goes beyond the aim of this work, what follows can be easily proved. Since the only source of the $A_4$ breaking is the VEVs of the flavons, any flavour structure is preserved above the corresponding energy scale, whatever interactions are present. In particular, the lagrangian (\ref{Ll}, \ref{Lnu}) contains all possible LO terms, given the group assignments, and its invariance under $A_4$ is maintained moving downward to the scale $\mean{\varphi}$, where significant changes in the flavour structure can appear.
From eqs.~(\ref{Y}) and (\ref{RHEigenvalues}), we deduce that $\mean{\varphi}\sim M_i$ and as a result in the AF model $\Delta_\tau$ must be proportional to $\ln(\mean{\varphi}/\lambda)$ and not to $\ln(\La_f/\lambda)$.

We will separately discuss the evolution of angles and phases for both type of hierarchy. In the following, the results will be shown for the SM and for the MSSM with $\tan\beta =15$ apart where explicitly indicated otherwise. Without loss of generality, we choose $y=1$ for our numerical analysis. We also set $\mean{\varphi}=10^{15}$.
The spectrum spans the range obtained in (\ref{RangeMasses}).

%%%%%%%%%%%%%%%%%%%%%%
%%%%%%%%%%%%%%%%%%%%%%

\subsection{Running of the angles}

\begin{figure}[ht!]
 \centering
\includegraphics[width=7.8cm]{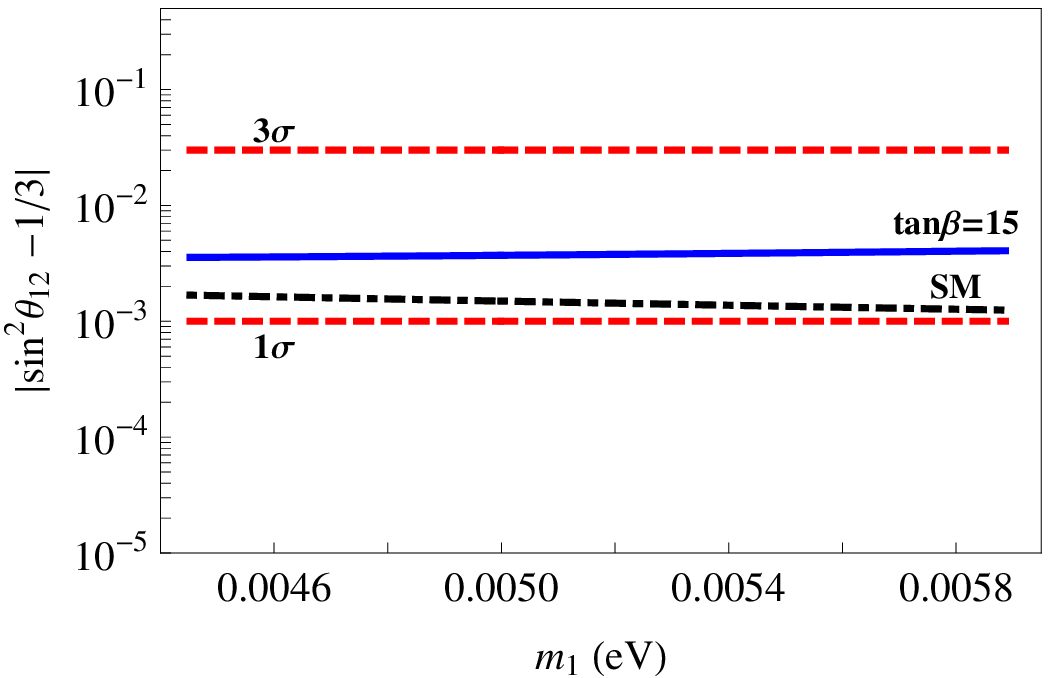}
\includegraphics[width=7.8cm]{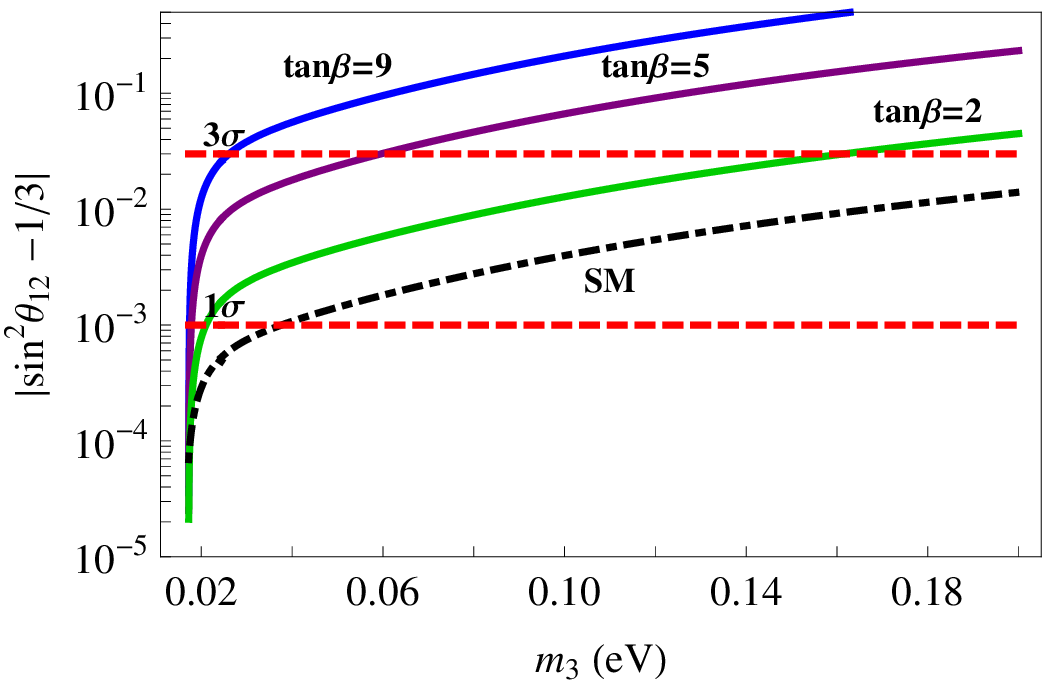}
\includegraphics[width=7.8cm]{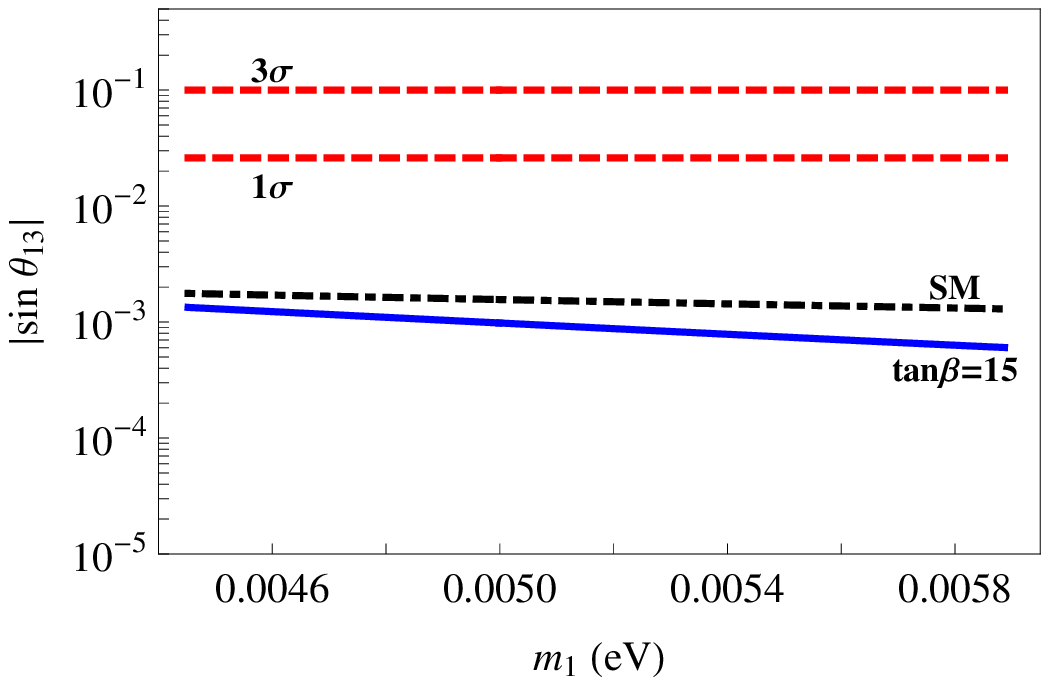}
\includegraphics[width=7.8cm]{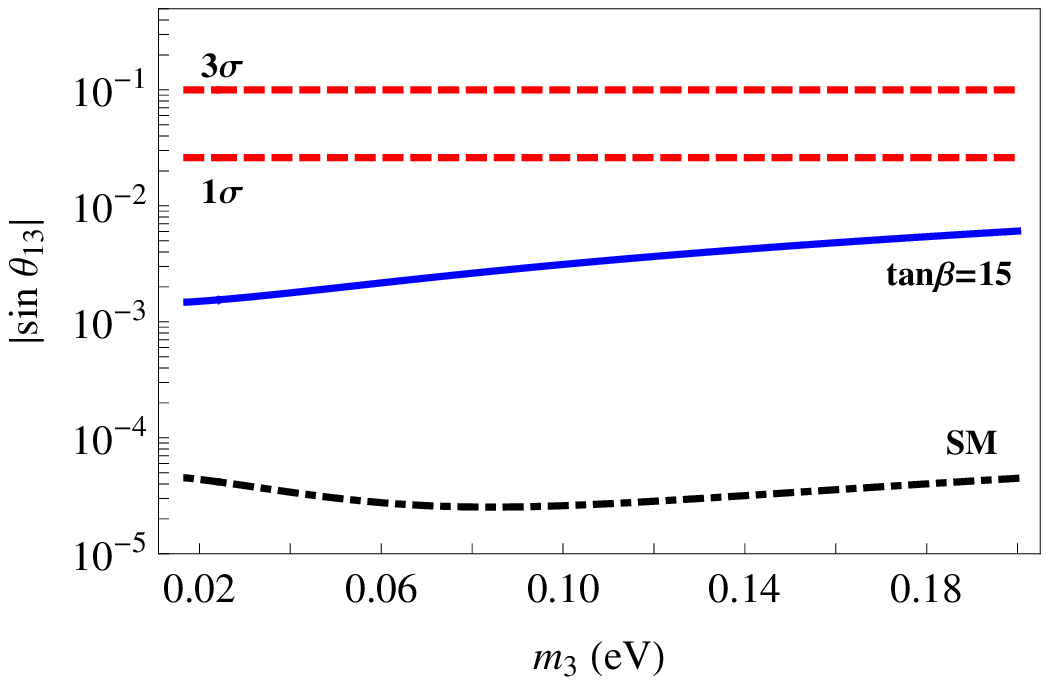}
\includegraphics[width=7.8cm]{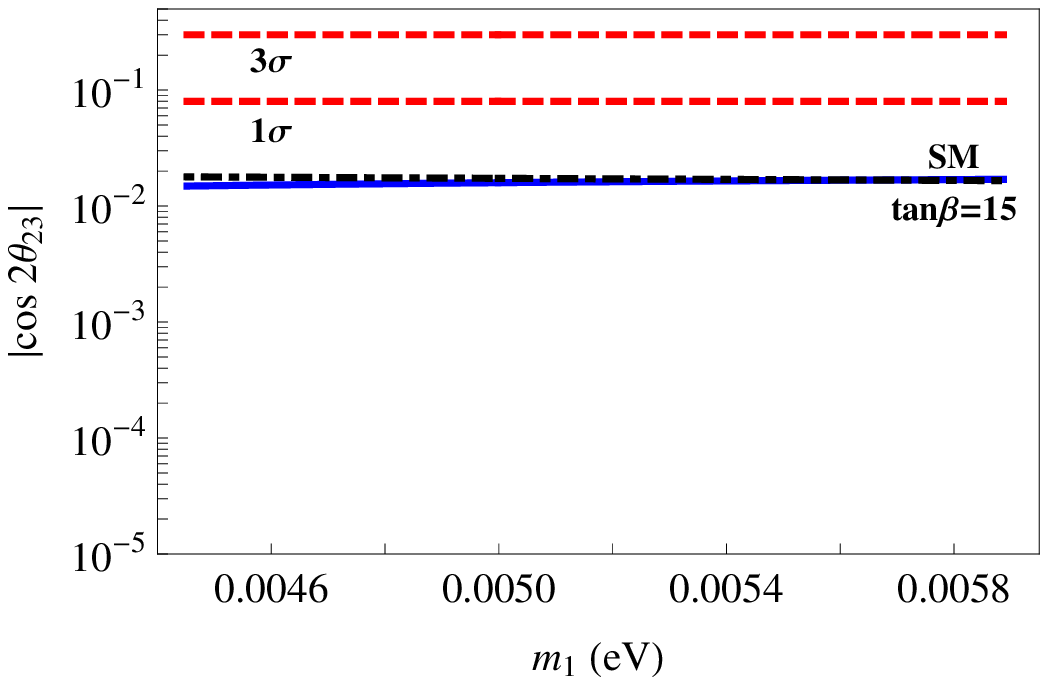}
\includegraphics[width=7.8cm]{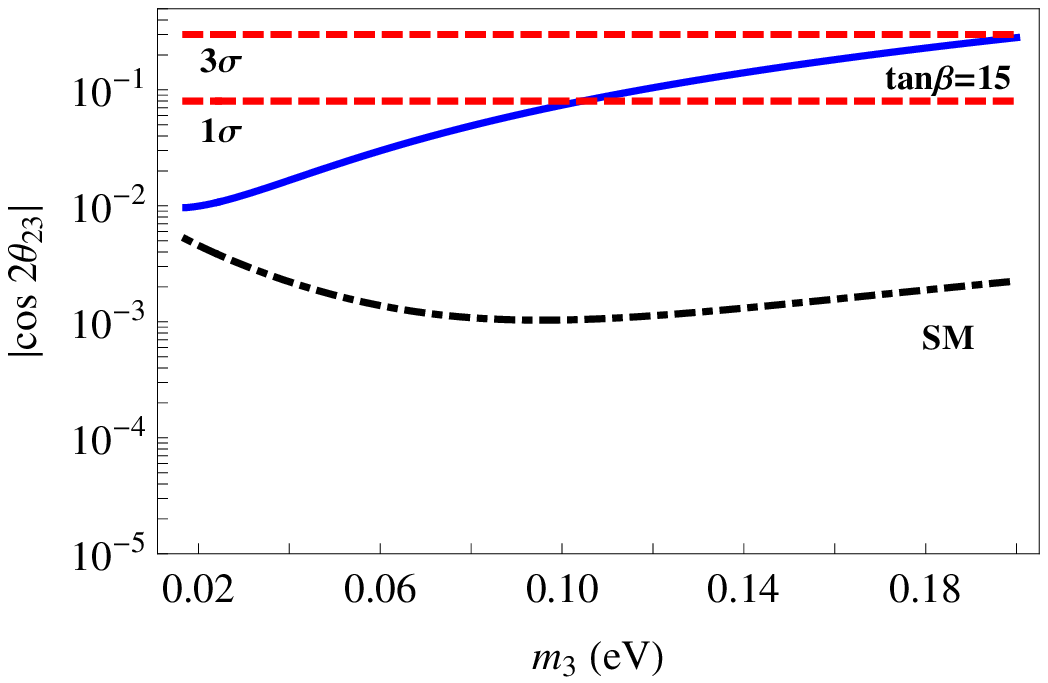}
\caption{Corrections to the TB mixing angles as functions of $m_1 (m_3)$ for the NH (IH) are shown. On the left column the plots refer to the NH spectrum, while on the right column the IH case is reported. The plots show the MSSM case with $\tan\beta =15$ (solid blue) and the SM case (black dashed), compared to the current $1\sigma$ and $3\sigma$ limits (dashed red). For the specific case of $|\sin^2\theta_{12}-1/3|$ for the IH, two values for $\tan\beta=2,\,5,\,9$ are considered (solid blue, purple and green). All the plots span in a range for $m_1 (m_3)$ which is given by eq. (\ref{RangeMasses}) or by the KATRIN bound.}
\label{fig:ANG}
\end{figure}

Since we are interested in deviations of the corrected mixing angles from the TB predictions
and in confronting them with experimental values, it is convenient
to relate the coefficients $k_{ij}$ defined in section (\ref{RGcoefficients}) with physical observables. Keeping in mind that $\vert k_{ij}\vert \ll 1$ and that we start from a TB mixing matrix, it follows that
\beq
\ba{ccc}
\sin\theta_{13}\simeq k_{13}\;,&\cos2 \theta_{23}\simeq 2 k_{23}\;,& \sin^2\theta_{12}-\dfrac{1}{3} \simeq \dfrac{2 \sqrt{2}}{3} k_{12}\;.
\ea
\eeq
The corrections to the TB mixing angles as functions of $m_1 (m_3)$ in the NH (IH) case are shown in figure \ref{fig:ANG}.

We begin with the case of NH.
Since the dependence of the corrected mixing angles from $\Delta_\tau$ is the same, SM corrections are generally expected to be smaller than those in MSSM. However, from figure \ref{fig:ANG}
we see that, in NH, there is not a large split between the two curves for SM and MSSM
respectively. This fact suggests a dominant contribution coming from the charged lepton sector. For the atmospheric and reactor angles,
the deviation from the TB prediction lays roughly one order of magnitude
below the $1\sigma$ limit. In particular, RG effects on $\sin\theta_{13}$ are even
smaller than the NLO contributions analyzed in section \ref{NLO} which
are of $\mathcal{O}(u)$, without cancellations.
On the other hand, since the experimental value of the solar angle is better measured than the other
two, the running effects become more important in this case.
Indeed, RG correction to the TB solar angle evades the $1\sigma$ limit as it can be clearly
seen in figure \ref{fig:ANG}.
Anyway, we observe that for both the atmospheric and solar angles, the running contribution is
of the same order as the contribution from NLO operators.

Now we move to analyze the case of IH.
In this case, since the neutrino spectrum predicted by the AF model
is almost degenerate, the contribution from the charged lepton sector in eqs. (\ref{EqDeltaYeDagYeIH}) is subdominant.
As a consequence the information which distinguishes the SM case from the MSSM one is mainly
dictated by $\Delta_\tau$ defined in eq.~(\ref{DeltaTau}).
As a result the running effects in the MSSM are always larger than in the SM
and for large $\tan \beta$ they are potentially dangerous.
The curves corresponding to the atmospheric and reactor angles
do not go above the $3\sigma$ and $1\sigma$ windows respectively.
However, the deviation from $\theta^{TB}_{12}$ presents a more interesting situation.
For example, for $\tan\beta \gtrsim 10$, the RG effects push
the value of the solar angle beyond the $3\sigma$ limit for entire spectrum.
For lower values of $\tan\beta$, the model is within the $3\sigma$ limit
only for a part of the spectrum where the neutrinos are less degenerate.
Confronting with the running effects, in the IH case, the contribution from NLO operators
in the AF model is under control.

%%%%%%%%%%%%%%%%%%%%%%
%%%%%%%%%%%%%%%%%%%%%%

\subsection{Running of the phases}
Majorana phases are affected by RG running effects too. Since there is not experimental information
on Majorana phases available in this moment we will simply show their values at low energy, eventually comparing them with the prediction in the AF model. We stress again that they are completely determined by only one parameter, the mass of the lightest neutrino, $m_1$ for NH and $m_3$ for IH.

\begin{figure}[h!]
 \centering
\includegraphics[width=7.8cm]{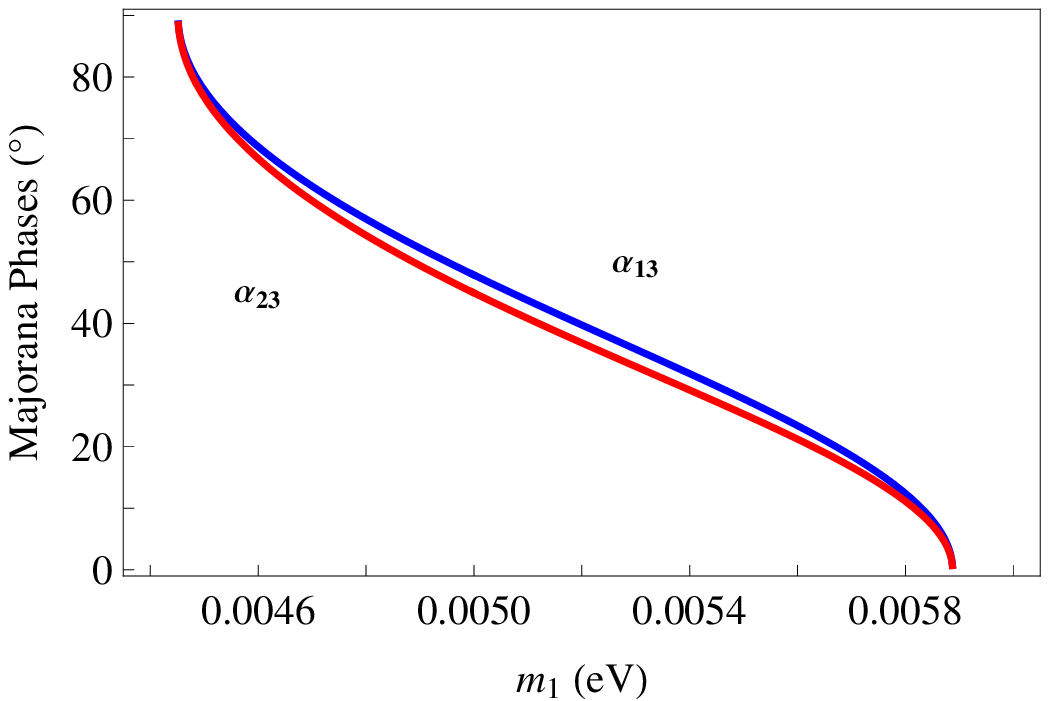}\quad
\includegraphics[width=7.8cm]{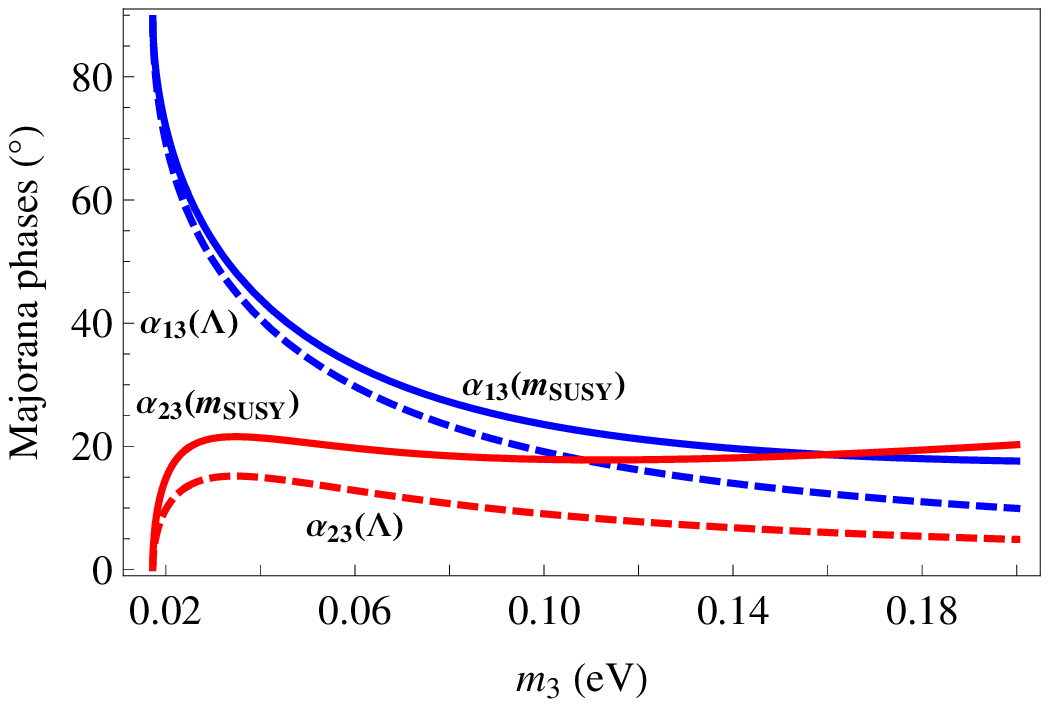}
\caption{Majorana phases $\alpha_{13}$ and $\alpha_{23}$ as functions of the lightest LH neutrino masses. For the NH (left panel) the corresponding curves at low and high energies are undistinguishable. For the IH (right panel) the curves refers to low energy values in MSSM with $\tan\beta =15$ (solid blue or red) and the AF prediction at $\Lambda_f$ (dashed blue or red).}
 \label{fig:MAJ}
\end{figure}

In the case of NH, Majorana phases are essentially not corrected by RG effects.
This feature is due to the fact that $\delta \alpha_{13}$ and
$\delta \alpha_{23}$ of eqs.~(\ref{Deltaalpha13}, \ref{Deltaalpha23}) are proportional to
$\sin(\alpha_{13}-\alpha_{23})$ which is close to zero, as we can see looking at the left panel of figure \ref{fig:MAJ}.
In the case of IH, MSSM RG effects
always increase the values of phases when moving from high energy
to low energy and they are maximized for $\tan\beta=15$, especially when the neutrino spectrum
becomes degenerate. On the contrary, in the SM context, the low energy curves
cannot be distinguished from the high energy ones.

\begin{figure}[h!]
 \centering
\includegraphics[width=7.8cm]{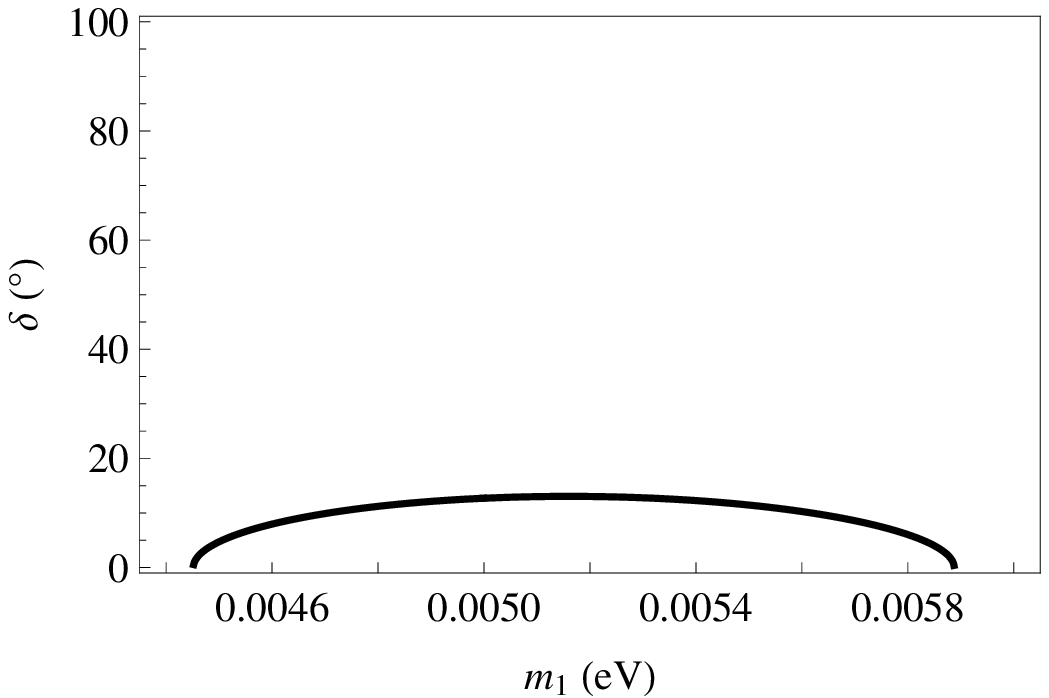}
\includegraphics[width=7.8cm]{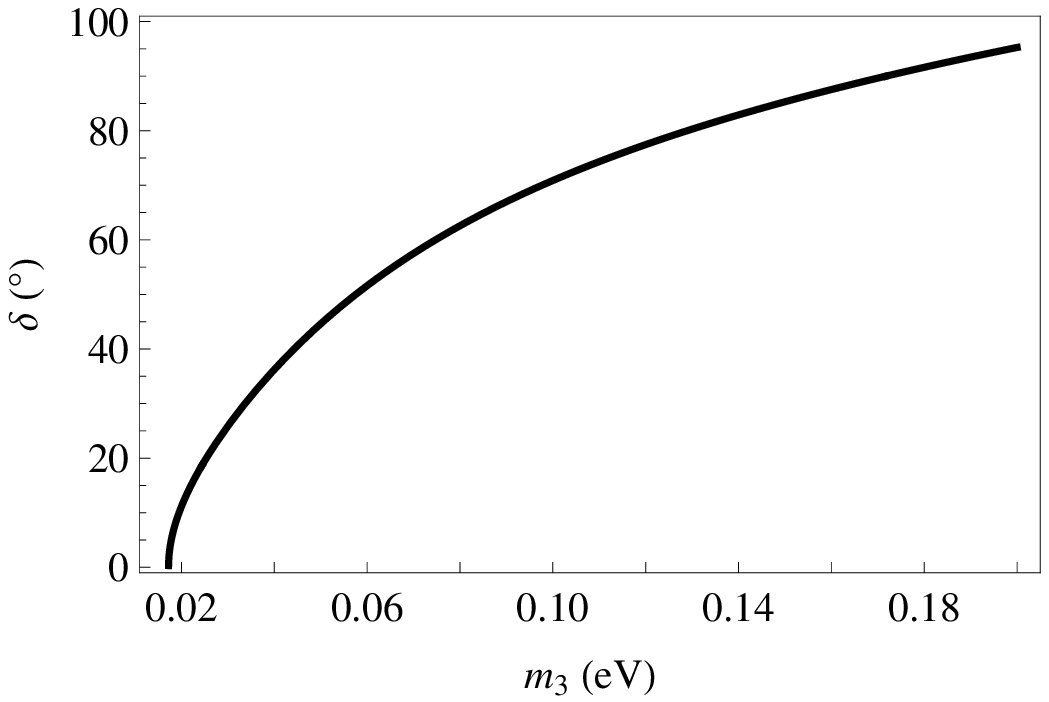}\\ [3mm]
\includegraphics[width=7.8cm]{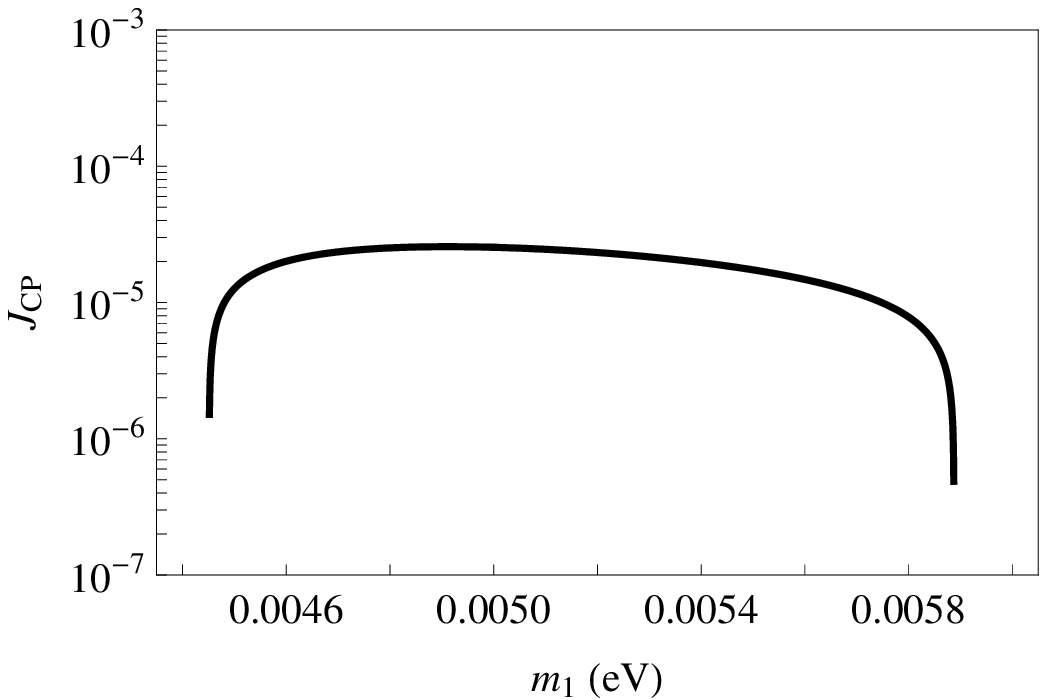}
\includegraphics[width=7.8cm]{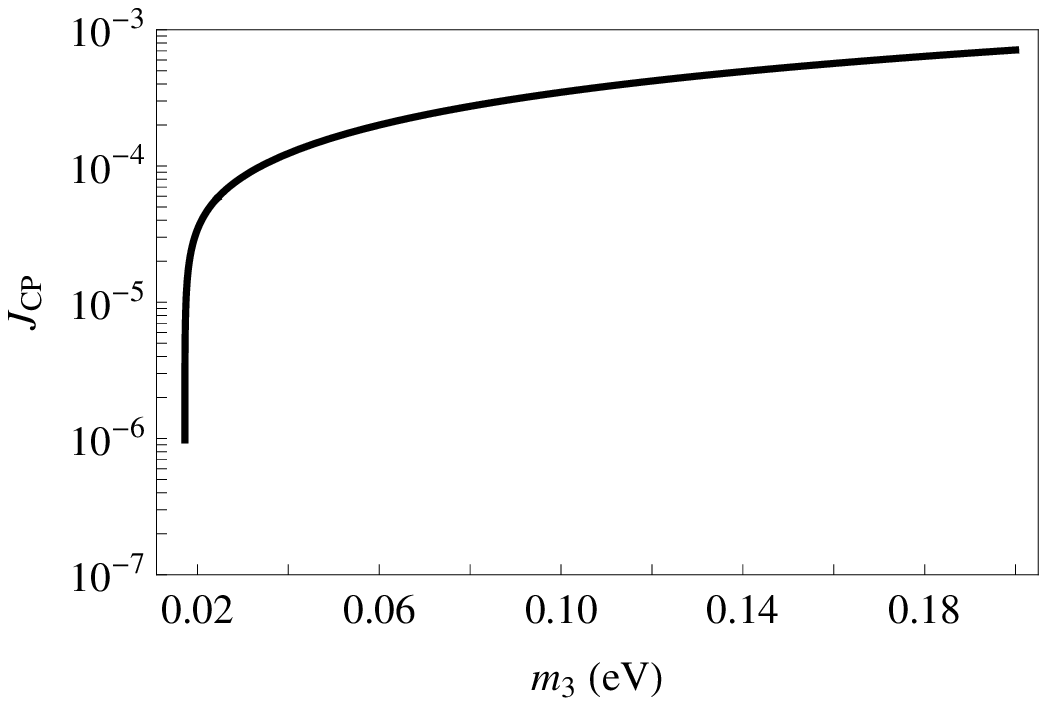}
\caption{Dirac CP phase $\delta_{CP}$ and Jarlskog invariant $J_{CP}$ as functions of the lightest LH neutrino masses for the NH (left column) and the IH (right column), in the MSSM with  $\tan\beta =15$.}
 \label{fig:JCP}
\end{figure}

As described in section (\ref{RGcoefficients}), a definite Dirac CP violating phase $\delta_{CP}$ arises from running effects even if, in the presence of a TB mixing pattern, it is undetermined in the beginning.
Although the final Dirac phase can be large, Jarlskog invariant,
which measures an observable CP violation, remains small because of the smallness of $\theta_{13}$. We remember that these results are valid both for the SM and for MSSM.

%%%%%%%%%%%%%%%%%%%%%%%%%%%%%%%%%%%%%%%%%%%%%%%%%%%%%%%%%%%%%%%%%%%%%%%%%%%%%%%%%%%%%%%%%%%%%%%%%%%%%%%%%%%%%%%%%%%%%%%%
%%%%%%%%%%%%%%%%%%%%%%%%%%%%%%%%%%%%%%%%%%%%%%%       5. CONCLUSIONS        %%%%%%%%%%%%%%%%%%%%%%%%%%%%%%%%%%%%%%%%%%%%
%%%%%%%%%%%%%%%%%%%%%%%%%%%%%%%%%%%%%%%%%%%%%%%%%%%%%%%%%%%%%%%%%%%%%%%%%%%%%%%%%%%%%%%%%%%%%%%%%%%%%%%%%%%%%%%%%%%%%%%%

\section{Conclusions and outlook}

The flavour sector is poorly known within the SM or the MSSM
and the masses and mixings are considered as
free parameters that can be adjusted to agree with experiments.
This lack can be improved by adding to the SM or the MSSM gauge group
appropriate flavour symmetries such that Yukawa couplings can be understood
in a more fundamental way. Even though the mechanism that generates
the fermion masses is not yet completely understood, a great effort has been
made in the last years, especially after the discovery of neutrino masses
which are significantly smaller than those of charged fermions.
Furthermore, the leptonic mixing pattern is also very different from $V_{\rm{CKM}}$
because it contains a nearly maximal atmospheric angle $\theta_{23}$ and
a solar angle very approximate to the TB prediction.
The flavour structure of the neutrino mass matrix can be nicely explained by (type I) seesaw mechanism implemented with flavour symmetries.
In order to naturally describe lepton masses and mixings, however,
the flavour group must be (spontaneously) broken.
Then we can expect that the LO prediction of a flavour symmetry
is always subject by subleading corrections characterized by small symmetry breaking parameters
($t$ and $u$ in our case). Then, in a consistent flavour model building, these subleading corrections must be safely under control.

In any flavour model, there is also an other independent correction to the LO predictions
which is due to the RG evolution of parameters. In this paper we have studied
these running effects on neutrino mixing patterns.
In seesaw models, the running contribution from the neutrino Yukawa coupling $Y_\nu$~,
encoded in $J_\nu$~, is generally dominant at energy above the seesaw threshold.
However, this effect, which in general introduces appreciable deviations from the LO mixing patterns,  does not affect the mixing angles, under specific conditions: in the first part of the paper, we have analyzed two classes of models in which this indeed happens.
The first class is characterized by a unitary $Y_\nu$~.
It is the case, for example, when the RH singlet neutrinos or the charged leptons being in a irreducible representation of the flavour group.
The second class is the so called mass-independent mixing pattern, in which,
in particular, the effect of $J_\nu$ can be absorbed by a small shift on neutrino mass
eigenvalues leaving mixing angles unchanged.
The widely studied TB mixing pattern belongs, for example, to this second class of models.

In the second part of the paper,  we focused on a
special realization of the general class of flavour models studied in the first part.
We were interested in the AF model for TB mixing
where the flavour symmetry group is given by $A_4 \times Z_3 \times U(1)_{\rm FN}$.
The aim is to analyze the RG effects on the TB mixing pattern in addition to
the NLO corrections already present in this model
and to confront them with experimental values.
The analysis has been performed both in the SM and MSSM and for both neutrino spectra.
We found that for NH light neutrinos, the dominant running contribution
comes from the charged lepton sector which weakly depends on both $\tan\beta$ and mass
degeneracy. As a result, for this type of spectrum, the tribimaximal prediction
is stable under RG evolution. Moreover, the running contribution is of the same order
or smaller with respect to the contribution from NLO operators.
On the other hand, in the case of IH, the deviation of the solar angle
from its TB value can be larger than the NLO contribution and, in particular,
for $\tan\beta \gtrsim 10$ an IH spectrum is strongly disfavored.
In the end, observe that for both spectra, the reactor angle $\theta_{13}$ does not receive
appreciable deviations from zero (at a level $\lesssim$ u).

The effects of RG running can be manifested also in other phenomena which are not directly related to the neutrino properties, such as Lepton Flavour Violating (LFV) transitions, leptogenesis, etc.
For example, in \cite{Lin_lepto, ABMMM_Lepto, BBFN_Lepto, TBlepto}, it has been pointed out that,
in the limit of an exact $A_4$ symmetry, all CP violating asymmetries vanish.
Then when the flavour symmetry is spontaneously broken, NLO corrections become important
in generating the desired leptogenesis. On the other hand, also the RG evolution can
introduce symmetry breaking effects which can be in principle
dominant over NLO contributions. As a result, the estimate of generated leptogenesis can be
quite different from those obtained in \cite{Lin_lepto, ABMMM_Lepto, BBFN_Lepto, TBlepto}
since they do not take into account the RG effects.
Another important consequence of RG running is the generation of off diagonal terms in the
soft SUSY breaking mass matrices, contributing to LFV rare processes. In a series of papers \cite{FHLM_LFVSM,FHLM_LFVMSSM,SUSYLFV+symmetries} it has been studied the impact of using flavour symmetries in order to explain the measured bounds on rare decays. In \cite{FHLM_LFVSM} it has already been shown that, below the seesaw scales, the running effect is negligible with respect to that originated in the
corresponding flavour theory. However between the seesaw scales the threshold effects are
 important and the two contributions to LFVs
can be comparable. All these issues are very interesting and are subject for a further investigation.

%%%%%%%%%%%%%%%%%%%%%%%%%%%%%%%%%%%%%%%%%%%%%%%%%%%%%%%%%%%%%%%%%%%%%%%%%%%%%%%%%%%%%%%%%%%%%%%%%%%%%%%%%%%%%%%%%%%%%%%%
%%%%%%%%%%%%%%%%%%%%%%%%%%%%%%%%%%%%%%%%%%%%%%%       AKNOWLEGMENTS        %%%%%%%%%%%%%%%%%%%%%%%%%%%%%%%%%%%%%%%%%%%%%
%%%%%%%%%%%%%%%%%%%%%%%%%%%%%%%%%%%%%%%%%%%%%%%%%%%%%%%%%%%%%%%%%%%%%%%%%%%%%%%%%%%%%%%%%%%%%%%%%%%%%%%%%%%%%%%%%%%%%%%%

\section*{Acknowledgments}
\addcontentsline{toc}{section}{Aknowledgements}

We thank Ferruccio Feruglio for many useful discussions. We thank Michael Schmidt for useful comments on the previous version of the paper.
We recognize that this work has been partly supported by the European Commission under contracts MRTN-CT-2006-035505 and by the European Programme ``Unification in the LHC Era'', contract PITN-GA-2009-237920 (UNILHC).
\vfill

\vfill
\newpage
%%%%%%%%%%%%%%%%%%%%%%%%%%%%%%%%%%%%%%%%%%%%%%%%%%%%%%%%%%%%%%%%%%%%%%%%%%%%%%%%%%%%%%%%%%%%%%%%%%%%%%%%%%%%%%%%%%%%%%%%
%%%%%%%%%%%%%%%%%%%%%%%%%%%%%%%%%%%%%%%%%%%%%%%       APPENDIX A       %%%%%%%%%%%%%%%%%%%%%%%%%%%%%%%%%%%%%%%%%%%%%%%%%
%%%%%%%%%%%%%%%%%%%%%%%%%%%%%%%%%%%%%%%%%%%%%%%%%%%%%%%%%%%%%%%%%%%%%%%%%%%%%%%%%%%%%%%%%%%%%%%%%%%%%%%%%%%%%%%%%%%%%%%%

\section*{Appendix A~: Renormalisation group equations}
\addcontentsline{toc}{section}{Appendix A}

In order to calculate the RG evolution of the light neutrino mass matrix, the RGEs for all the parameters of the theory have to be solved simultaneously. We use the notation defined in the text, where a superscript $(n)$ denotes a quantity between the $n$th and the $(n+1)$th mass threshold. When all the RH neutrinos are integrated out, the RGEs can be recovered by setting the neutrino Yukawa coupling $Y_\nu$ to zero, while in the full theory above the highest seesaw scale, the superscript $(n)$ has to be omitted.

In the SM extended by singlet neutrinos, the RGEs for ${Y}_e$, $\accentset{(n)}{Y}_\nu$, $\accentset{(n)}{M}$, $\accentset{(n)}{\kappa}$, ${Y}_d$, and ${Y}_u$ are given by
\beq
\ba{ccl}
16\pi^2 \accentset{(n)}{Y_e} & = & Y_e \left\{ \dfrac{3}{2} Y_e^\dagger Y_e -\dfrac{3}{2} \accentset{(n)}{Y}_\nu^\dagger \accentset{(n)}{Y}_\nu +\Tr\left[ 3\,Y_u^\dagger Y_u + 3\,Y_d^\dagger Y_d + \accentset{(n)}{Y}_\nu^\dagger \accentset{(n)}{Y}_\nu + Y_e^\dagger Y_e\right] -\dfrac{9}{4} g_1^2 - \dfrac{9}{4} g_2^2\right\}\;,\nn\\
&&\nn\\
16\pi^2 \accentset{(n)}{Y_\nu} & = & \accentset{(n)}{Y}_\nu \left\{ \dfrac{3}{2} \accentset{(n)}{Y}_\nu^\dagger \accentset{(n)}{Y}_\nu - \dfrac{3}{2} Y_e^\dagger Y_e +\Tr\left[ 3\,Y_u^\dagger Y_u + 3\,Y_d^\dagger Y_d + \accentset{(n)}{Y}_\nu^\dagger \accentset{(n)}{Y}_\nu + Y_e^\dagger Y_e\right] -\dfrac{9}{20} g_1^2 -\dfrac{9}{4} g_2^2 \right\}\;,\nn\\
&&\nn\\
16\pi^2 \accentset{(n)}{M} &=& \Big(\accentset{(n)}{Y}_\nu \accentset{(n)}{Y}^\dagger_\nu\Big) \accentset{(n)}{M} + \accentset{(n)}{M} \Big(\accentset{(n)}{Y}_\nu \accentset{(n)}{Y}^\dagger_\nu\Big)^T \;,\nn\\
&&\nn\\
16\pi^2\accentset{(n)}{\kappa} & = & \dfrac{1}{2} \Big[ \accentset{(n)}{Y}^\dagger_\nu \accentset{(n)}{Y}_\nu -3 Y_e^\dagger Y_e \Big]^T \accentset{(n)}{\kappa} +\dfrac{1}{2}\accentset{(n)}{\kappa} \Big[\accentset{(n)}{Y}^\dagger_\nu \accentset{(n)}{Y}_\nu -3 Y_e^\dagger Y_e \Big]+\nn\\[3mm]
&&+2 \Tr\left[ 3\,Y_u^\dagger Y_u + 3\,Y_d^\dagger Y_d + \accentset{(n)}{Y}_\nu^\dagger \accentset{(n)}{Y}_\nu + Y_e^\dagger Y_e\right]
- 3 g_2^2\accentset{(n)}{\kappa} +\lambda_H\accentset{(n)}{\kappa}\;,\nn\\
&&\nn\\
16\pi^2 \, \accentset{(n)}{Y_d} & = & Y_d \left\{ \dfrac{3}{2} Y_d^\dagger Y_d - \dfrac{3}{2} Y_u^\dagger Y_u + \Tr\Big[3\,Y_u^\dagger Y_u + 3\,Y_d^\dagger Y_d + \accentset{(n)}{Y}_\nu^\dagger \accentset{(n)}{Y}_\nu + Y_e^\dagger Y_e \Big] - \dfrac{1}{4} g_1^2 - \frac{9}{4} g_2^2 - 8g_3^2\right\}\;,\nn\\
&&\nn\\
16\pi^2 \, \accentset{(n)}{Y_u} & = & Y_u \left\{ \dfrac{3}{2} Y_u^\dagger Y_u - \dfrac{3}{2} Y_d^\dagger Y_d + \Tr\Big[3\,Y_u^\dagger Y_u + 3\,Y_d^\dagger Y_d + \accentset{(n)}{Y}_\nu^\dagger \accentset{(n)}{Y}_\nu + Y_e^\dagger Y_e \Big] - \dfrac{17}{20} g_1^2 - \dfrac{9}{4} g_2^2 - 8g_3^2 \right\}\;,\nn\\
&&\nn\\
16\pi^2\,\accentset{(n)}{\lambda_H} & = & 6\lambda_H^2 -3\lambda_H \left(3g_2^2+\dfrac{3}{5} g_1^2\right) +3 g_2^4 +\dfrac{3}{2}\left(\dfrac{3}{5} g_1^2+g_2^2\right)^2 +\\[3mm]
&&+4\lambda_H \Tr\Big[3\,Y_u^\dagger Y_u + 3\,Y_d^\dagger Y_d + \accentset{(n)}{Y}_\nu^\dagger \accentset{(n)}{Y}_\nu + Y_e^\dagger Y_e \Big]+\nn\\[3mm]
&&-8 \Tr\Big[ 3\,Y_u^\dagger Y_u\,Y_u^\dagger Y_u +  3\,Y_d^\dagger Y_d\,Y_d^\dagger Y_d + \accentset{(n)}{Y}_\nu^\dagger\accentset{(n)}{Y}_\nu \accentset{(n)}{Y}_\nu^\dagger\accentset{(n)}{Y}_\nu + Y_e^\dagger Y_e\,Y_e^\dagger Y_e \Big]\;.\nn
\ea
\eeq
We use the convention that the Higgs self-interaction term in the
Lagrangian is $-\lambda_H (H^\dagger H)^2/4$.

In the MSSM context the 1-loop RGEs for the same quantities are given by
\beq
\ba{ccl}
16\pi^2 \dfrac{d}{d t}\accentset{(n)}{Y_e} & = & Y_e\left\{ 3Y_e^\dagger Y_e +\accentset{(n)}{Y}_\nu ^\dagger  \accentset{(n)}{Y}_\nu + \Tr\Big[3Y_d^\dagger Y_d +Y_e^\dagger Y_e\Big] - \frac{9}{5}g_1^2 - 3g_2^2\right\}\;,\\
\nn\\
16\pi^2 \dfrac{d}{d t}\accentset{(n)}{Y_\nu} &=& \accentset{(n)}{Y}_\nu \left\{ 3 \accentset{(n)}{Y}^\dagger_\nu \accentset{(n)}{Y}_\nu + Y_e^\dagger Y_e + \Tr\Big[3Y_u^\dagger Y_u + \accentset{(n)}{Y}^{\dagger}_\nu \accentset{(n)}{Y}_\nu\Big] - \dfrac{3}{5} g_1^2 - 3 g_2^2 \right\}\;,\\
\nn\\
16\pi^2 \dfrac{d}{d t} \accentset{(n)}{M}_R &=& \vphantom{\dfrac{1}{2}} 2\Big(\accentset{(n)}{Y}_\nu \accentset{(n)}{Y}^\dagger_\nu\Big) \accentset{(n)}{M}_R + 2\accentset{(n)}{M}_R\Big(\accentset{(n)}{Y}_\nu \accentset{(n)}{Y}^\dagger_\nu\Big)^T\;,\\
\nn\\
16\pi^2 \dfrac{d}{d t}\accentset{(n)}{\kappa} & = & \Big[\accentset{(n)}{Y}^\dagger_\nu \accentset{(n)}{Y}_\nu + Y_e^\dagger Y_e \Big]^T \accentset{(n)}{\kappa} + \accentset{(n)}{\kappa} \Big[\accentset{(n)}{Y}^\dagger_\nu\accentset{(n)}{Y}_\nu + Y_e^\dagger Y_e \Big] + 2 \Tr\Big[3 Y_u^\dagger Y_u + \accentset{(n)}{Y}^{\dagger}_\nu \accentset{(n)}{Y}_\nu \Big]\accentset{(n)}{\kappa} -\frac{6}{5} g_1^2 \accentset{(n)}{\kappa}- 6 g_2^2 \accentset{(n)}{\kappa}\;,\\
\nn\\
16\pi^2 \dfrac{d}{d t}\accentset{(n)}{Y_d}& = & Y_d\left\{ 3Y_d^\dagger Y_d + Y_u^\dagger Y_u + \Tr\Big[3 Y_d^\dagger Y_d + Y_e^\dagger Y_e\Big] - \dfrac{7}{15}g_1^2 - 3g_2^2 - \dfrac{16}{3}g_3^2 \right\}\;,\\
\nn\\
16\pi^2 \dfrac{d}{d t} \accentset{(n)}{Y_u} & = & Y_u\left\{ Y_d^\dagger Y_d + 3 Y_u^\dagger Y_u + \Tr\Big[3Y_u^\dagger Y_u + \accentset{(n)}{Y}_\nu ^\dagger  \accentset{(n)}{Y}_\nu\Big] - \dfrac{13}{15}g_1^2- 3g_2^2 - \dfrac{16}{3}g_3^2\right\}\;.
\ea\nn
\eeq

\vfill
\newpage
%%%%%%%%%%%%%%%%%%%%%%%%%%%%%%%%%%%%%%%%%%%%%%%%%%%%%%%%%%%%%%%%%%%%%%%%%%%%%%%%%%%%%%%%%%%%%%%%%%%%%%%%%%%%%%%%%%%%%%%%
%%%%%%%%%%%%%%%%%%%%%%%%%%%%%%%%%%%%%%%%%%%%%%%       BIBLIOGRAPHY        %%%%%%%%%%%%%%%%%%%%%%%%%%%%%%%%%%%%%%%%%%%%%%
%%%%%%%%%%%%%%%%%%%%%%%%%%%%%%%%%%%%%%%%%%%%%%%%%%%%%%%%%%%%%%%%%%%%%%%%%%%%%%%%%%%%%%%%%%%%%%%%%%%%%%%%%%%%%%%%%%%%%%%%

\end{document}